\documentclass[journal,twoside,web]{ieeecolor}

\usepackage{generic}
\usepackage{cite}
\usepackage{amsmath}
\usepackage{amssymb}
\usepackage{amsfonts}
\usepackage{graphicx}
\usepackage{enumerate}
\usepackage{algorithm}
\usepackage{algpseudocode}
\usepackage{mathrsfs}
\usepackage{float}
\usepackage{tikz}
\usetikzlibrary{matrix,arrows,decorations.pathmorphing}
\usepackage{verbatim}
\usepackage{mathtools}
\usepackage{caption}
\usepackage{subcaption}
\usetikzlibrary{arrows}
\usepackage{tabularx}

\newtheorem{assumption}{Assumption}
\newtheorem{corollary}{Corollary}

\algrenewcommand\algorithmicrequire{\textbf{Input:}}
\algrenewcommand\algorithmicensure{\textbf{Output:}}

\newcommand{\hma}[1]{\textcolor{black}{#1}}

\graphicspath{{./figs/}}

\def\BibTeX{{\rm B\kern-.05em{\sc i\kern-.025em b}\kern-.08em
    T\kern-.1667em\lower.7ex\hbox{E}\kern-.125emX}}

\begin{document}

\title{Tensor-Algebraic Framework for Nonlinear Control: Spectral Stability Analysis and Controller Design of Polynomial Systems}

\author{Shaoxuan Cui$^{1}$, Guofeng Zhang$^{1}$, Hildeberto Jardón-Kojakhmetov$^{2}$, and Ming Cao$^{3}$
\thanks{$^{1}$ S. Cui and G. Zhang are with the Department of Applied Mathematics, The Hong Kong Polytechnic University, Kowloon 999077, Hong Kong, China.
(e-mail: \{shaoxuan.cui, guofeng.zhang\}@polyu.edu.hk).}
\thanks{$^{2}$ H. Jard\'on-Kojakhmetov is with the Bernoulli Institute for Mathematics, Computer Science and Artificial Intelligence, University of Groningen, Groningen, 9747 AG Netherlands
(e-mail: h.jardon.kojakhmetov@rug.nl).}
\thanks{$^{3}$ M. Cao is with the Engineering and Technology institute Groningen, University of Groningen, Groningen, 9747 AG Netherlands
(e-mail: m.cao@rug.nl).}
}

\maketitle

\newtheorem{remark}{Remark}
\newtheorem{lemma}{Lemma}
\newtheorem{thm}{Theorem}
\newtheorem{example}{Example}
\newtheorem{definition}{Definition}
\newtheorem{prop}{Proposition}


\begin{abstract}
Matrices provide an algebraic foundation for systems and control, whereas an analogous tensor-based framework for general nonhomogeneous polynomial dynamics is still lacking. This paper addresses this gap by developing a tensor-algebraic framework that connects tensor structure, spectral properties, Lyapunov stability, and controller synthesis. We first establish stability conditions based on positive-definite, copositive, and $P$-tensors for homogeneous and nonhomogeneous polynomial systems. We then introduce an extended tensor-eigenvalue problem (ETEP) for general multivariate polynomials, yielding a spectral--variational characterization of positivity through one interior and two boundary spectral quantities, and consequently a stability test for polynomial vector fields. For controller synthesis, we develop an unfolding-based U-eigenvalue relaxation that combines a positive-definite anchor with a near-positive-semidefinite higher-order part, leading to a convex semidefinite program. The conservatism of the relaxation is explicitly characterized, and an explicit guaranteed domain-of-attraction bound is derived. Numerical examples demonstrate positivity certification beyond direct SOS representations and substantially lower computation time than SOS-based synthesis in the considered example. These results establish a systematic tensor-spectral methodology for the analysis and control of polynomial systems with higher-order interactions.
\end{abstract}


\section{Introduction}

\IEEEPARstart{M}atrices provide the algebraic foundation of classical systems and control theory, where structural and spectral properties are closely connected to stability, positivity, and controller design~\cite{khalil2002nonlinear,wonham1974linear,FB-LNS}. For instance, matrix positivity, Metzler structures, and Lyapunov inequalities lead to tractable conditions for analyzing dynamical systems~\cite{narendra2010hurwitz,gurvits2007stability,cheng2003quadratic}. This has established a coherent framework linking algebraic properties of matrices with the qualitative behavior of the systems they represent.

Many modern systems, however, involve \emph{higher-order interactions} that cannot be adequately described by pairwise matrix couplings~\cite{bick2023higher}. Examples arise in social and spreading processes, biochemical and ecological networks, and multi-agent systems, where the evolution of an agent may depend simultaneously on several other components~\cite{cui2023general,cui2025analysis,cui2025higher}. Such interactions naturally lead to multilinear or polynomial dynamical systems whose coefficients are more conveniently represented by tensors. In parallel, tensor algebra has developed extensions of classical matrix notions, including eigenvalues, positive definiteness, copositivity, and diagonal dominance~\cite{qit,chang2013survey,wang2021hypergraph}. These developments provide a rich algebraic foundation, but their systematic use in systems and control remains comparatively limited.

Existing tensor-based results in control and dynamical systems have mainly focused on homogeneous or specially structured settings. Tensor eigenvalues have been used to test positive definiteness of homogeneous multivariate forms, with direct motivation from Lyapunov stability analysis~\cite{ni2008eigenvalue}. For homogeneous polynomial dynamical systems, orthogonally decomposable tensors enable explicit trajectory representations and sharp stability characterizations through tensor eigenpairs~\cite{chen2022explicit}. Tensor-based models have also been studied in application-specific settings such as hypergraph diffusion and higher-order Lotka--Volterra dynamics~\cite{cui2023general,cui2025analysis}. These results demonstrate the potential of tensor methods for system analysis and control, but they largely rely on homogeneity, particular tensor decompositions, or problem-specific structures. A general tensor-spectral framework that accommodates nonhomogeneous polynomial dynamics and further supports tractable controller synthesis is still lacking.

In this paper, we consider general tensor polynomial systems of the form
\[
\dot{x}=A_{k-1}x^{k-1}+A_{k-2}x^{k-2}+\cdots+A_1x,
\]
where the coefficient tensors may have different orders and no special decomposition or structure is assumed. Unlike homogeneous tensor systems, the simultaneous presence of terms of different degrees prevents their stability from being characterized directly by the spectrum of a single tensor. In particular, a systematic spectral characterization of nonhomogeneous polynomial positivity, together with a computationally tractable connection to Lyapunov stability and feedback design, is still missing. Addressing this gap requires extending tensor spectral analysis beyond homogeneous forms while retaining a clear connection between algebraic tensor properties and nonlinear system behavior.

\noindent\textbf{Contributions.}
This paper develops a unified tensor-based framework for the analysis and control of polynomial dynamical systems. First, in Sections III--VI, we derive algebraic conditions for positivity and stability of multilinear and polynomial tensor systems, and connect positive-definite, copositive, and $P$-tensors with Lyapunov stability. Second, in Section VII, we establish an extended tensor--eigenvalue problem (ETEP) for general nonhomogeneous polynomials, yielding a spectral--variational positivity criterion through interior and boundary spectral quantities and a corresponding stability test for polynomial vector fields. Finally, in Sections VIII--IX, we develop an unfolding-based U-eigenvalue relaxation for controller synthesis, characterize its conservatism, and propose a convex SDP combining a low-order positive-definite anchor with a near-PSD higher-order part. An explicit guaranteed domain-of-attraction bound is further derived, providing a computationally efficient alternative to SOS-based synthesis.

Together, these results extend classical system theory from matrices to tensors and establish 
a comprehensive tensor-based framework for nonlinear systems with higher-order interactions. 
They bridge tensor algebra, spectral theory, and nonlinear control, 
offering new perspectives on the analysis, stabilization, and spectral characterization of polynomial dynamical systems.

\emph{Notation:} Throughout the paper, $\mathbb{R}$ denotes the real numbers, 
$\mathbb{R}_{+}$ and $\mathbb{R}_{++}$ the nonnegative and positive reals, respectively; 
$\mathbb{R}^n$ the $n$-dimensional real space; 
$\mathbb{R}^{[n,m]}$ the space of $m$th-order $n$-dimensional real tensors; 
$\mathbb{R}^n_{+}$ and $\mathbb{R}^n_{++}$ the nonnegative and positive orthants in $\mathbb{R}^n$; 
$\mathbb{S}^{n-1}=\{u\in\mathbb{R}^n:\|u\|=1\}$ the unit sphere; 
and index subsets such as $I\subset\{1,\ldots,n\}$ are used where appropriate.
For $M \in \mathbb{R}^{n \times n}$, $\rho(M)$ is its spectral radius.
For $M \in \mathbb{R}^{n \times r}$ and $a \in \mathbb{R}^n$, $M_{ij}$ and $a_i$ are the $(i,j)$-th entry of $M$ and $i$-th entry of $a$.
For $a,b \in \mathbb{R}^n$, $a>(<)b$ means $a_i>(<)b_i$ for all $i$, and $a\geq(\leq)b$ similarly.
These comparisons extend to matrices or tensors of the same size.
$\mathbf{1}$ and $\mathbf{0}$ denote all-ones and all-zeros with appropriate dimensions.
Additional tensor-related preliminaries appear in the next section.

\section{Preliminaries}

A real tensor $T \in \mathbb{R}^{n_1 \times n_2 \times \cdots \times n_m}$ is a multidimensional array. 
The \emph{order} of $T$ is $m$, representing the number of its modes, and each dimension $n_i$, for $i = 1, \ldots, m$, is referred to as a \emph{mode}. 
If all modes have the same dimension $n$, the tensor is called \emph{cubical} and is denoted by $T \in \mathbb{R}^{[n,m]}$. 
A cubical tensor $T$ is said to be \emph{supersymmetric} if its entries remain invariant under any permutation of indices.

The \emph{identity tensor}, denoted by $\mathcal{I}=\{\delta_{i_1 \cdots i_m}\}$, is defined elementwise as
\begin{equation*}
   \delta_{i_1 \cdots i_m} = 
   \begin{cases}
   1, & \text{if } i_1 = i_2 = \cdots = i_m, \\
   0, & \text{otherwise.}
   \end{cases}
\end{equation*}

Throughout this paper, unless otherwise stated, the term ``tensor'' refers to a cubical tensor. 
We now recall several types of tensor eigenvalue problems that will be used in the sequel.

Consider a tensor $A \in \mathbb{R}^{[n,m]}$. 
The \emph{H-eigenvalue problem} \cite{qit} is defined by
\begin{equation}\label{eq:eigenproblem}
A x^{m-1} = \lambda x^{[m-1]},
\end{equation}
where $A x^{m-1}$ and $x^{[m-1]}$ are vectors whose $i$th components are
\begin{align*}
\left(A x^{m-1}\right)_i & = \sum_{i_2, \ldots, i_m=1}^n A_{i,i_2 \cdots i_m} x_{i_2} \cdots x_{i_m}, \\
\left(x^{[m-1]}\right)_i & = x_i^{m-1}.
\end{align*}
If there exist a real scalar $\lambda$ and a nonzero real vector $x$ satisfying~\eqref{eq:eigenproblem}, 
then $\lambda$ is called an \emph{H-eigenvalue} of $A$ and $x$ the associated \emph{H-eigenvector}. 
If $x > \mathbf{0}$ (i.e., all entries are strictly positive), then $\lambda$ and $x$ are referred to as an $H^{++}$-eigenvalue and $H^{++}$-eigenvector, respectively.
In this paper we explicitly specify the type of eigenpair (H-, Z-, or E-eigenpair) whenever needed to avoid ambiguity.

In this paper, we also use the following notation: for $A \in \mathbb{R}^{[n,m]}$,
\begin{equation*}
    Ax^m=x^\top Ax^{m-1}=\sum_{i_1, \ldots, i_m=1}^n A_{i_1,i_2 \cdots i_m} x_{i_1} \cdots x_{i_m}.
\end{equation*}

Another commonly used tensor eigenvalue concept is the \emph{Z-eigenvalue}~\cite{qit,chang2013survey}. 
For a tensor $A\in\mathbb{R}^{[n,m]}$, the pair $(\lambda, x)$ satisfies
\begin{equation}\label{eq:zeigenproblem}
    A x^{m-1}=\lambda x, \qquad x^\top x=1.
\end{equation}
If both $\lambda$ and $x$ are real, they are called a \emph{Z-eigenvalue} and \emph{Z-eigenvector} of $A$; 
if they are complex, they are referred to as an \emph{E-eigenvalue} and \emph{E-eigenvector}, respectively~\cite{qit,chang2013survey}.


A tensor $\mathcal{C} = (c_{i_1 \cdots i_m})\in \mathbb{R}^{[n,m]}$ is called \emph{reducible} 
if there exists a nonempty proper subset $I \subset \{1, \ldots, n\}$ such that
$c_{i_1 \cdots i_m}=0, \quad \forall i_1 \in I, \ \forall i_2, \ldots, i_m \notin I.$
If no such subset exists, $\mathcal{C}$ is said to be \emph{irreducible}. 
A tensor with all nonnegative entries is called a \emph{nonnegative tensor}.

For a matrix $R \in \mathbb{R}^{n \times n}$ and a tensor $A \in \mathbb{R}^{[n,m]}$, 
their product is defined as
$(RA)_{i_1 i_2 \ldots i_m} = \sum_{j=1}^{n} R_{i_1 j} A_{j i_2 \ldots i_m}.$

Following~\cite{chen2022explicit}, an $m$th-order $n$-dimensional tensor $A\in\mathbb{R}^{[n,m]}$ 
is said to be \emph{almost symmetric} if it is symmetric with respect to its last $m-1$ indices. 
For example, a third-order tensor $A \in \mathbb{R}^{n\times n\times n}$ is almost symmetric if
\[
A_{i_1i_2i_3}=A_{i_1i_3i_2}, \quad 
A_{i_2i_1i_3}=A_{i_2i_3i_1}, \quad 
A_{i_3i_1i_2}=A_{i_3i_2i_1}.
\]
Note that \cite{chen2022explicit} defines almost symmetry with respect to the \emph{first} $m-1$ indices, 
whereas here we adopt symmetry with respect to the \emph{last} $m-1$ indices. 
These two definitions are equivalent, and the difference arises from the tensor–vector multiplication convention. 
The present notation aligns with the one used in tensor algebra~\cite{qit,qi2013symmetric}.

\begin{lemma}\cite{chen2022explicit}\label{lemma_1}
Let $f:\mathbb{R}^n\to\mathbb{R}$ be a homogeneous scalar polynomial of degree $m$. 
Then there exists a unique supersymmetric tensor $A$ of order $m$ such that $f(x)=A x^{m}$. 
Similarly, for a homogeneous vector polynomial of degree $m-1:$ $g:\mathbb{R}^n\to\mathbb{R}^n$, 
there exists a unique almost symmetric tensor $A$ of order $m$ such that $g(x)=A x^{m-1}$.
\end{lemma}

It follows directly that if $A$ is supersymmetric, then the derivative of $f(x)=A x^m$ is
$\frac{df(x)}{dx}=m A x^{m-1}.$
If $A$ is almost symmetric, then the Jacobian of the vector field $g(x)=A x^{m-1}$ is
$\frac{\partial g(x)}{\partial x}=(m-1)A x^{m-2}.$

\section{Positive Tensor-Based Polynomial System}\label{sec:positive_tensor_system}

In this section, we investigate the positivity properties of tensor-based dynamical systems. Recall that a dynamical system $\dot{x}=f(x)$ is said to be \emph{positive} if, for any initial condition $x(0) \ge \mathbf{0}$, its trajectory $x(t)$ remains nonnegative for all $t\ge0$. Positivity is an essential property for systems representing population dynamics \cite{cui2025analysis}, opinion formation \cite{cui2023general}, or other physical processes in which states cannot become negative. We extend the classical theory of positive systems to multilinear tensor-based dynamics. Consider a general multilinear system
\begin{equation}\label{eq:multi_tensor_system}
    \dot{x} = - A x^{k-1},
\end{equation}
where $A \in \mathbb{R}^{[n,k]}$ is a $k$th-order $n$-dimensional tensor.
Throughout this section, we assume $x(0)\geq \mathbf{0}$ and study sufficient and necessary conditions for the system to preserve nonnegativity.

\subsection{Basic Sufficient Conditions}
Define the multi-index set $I=i_2,\cdots,i_k$, where $k$ is the order of $A$.
\begin{lemma}\label{lem:sufficient_positive}
Consider the multilinear tensor-based dynamical system \eqref{eq:multi_tensor_system}. Suppose that for any nonzero entry $A_{i,I}>0$, the index $i$ belongs to the multi-index set $I$. Then, the system is positive.
\end{lemma}
\begin{proof}
Let $x(0)\ge0$, and let $i$ be an index such that $x_i=0$. We have
   $ \dot{x}_i = - \sum_{A_{i,I}>0} A_{i,I} \prod_{j\in I} x_j - \sum_{A_{i,I}\le0} A_{i,I} \prod_{j\in I} x_j.$
As $x_i=0$, all products in the first sum vanish, and the remaining terms satisfy $\dot{x}_i \ge0$. Hence, the vector field points inward or along the boundary of the nonnegative orthant, implying that the system is positive.
\end{proof}

The above lemma provides a simple structural condition. It requires that positive tensor entries contribute only when the affected component appears in the corresponding interaction set.

\begin{lemma}\label{lem:nonhomogeneous_positive}
Consider a non-homogeneous tensor-based dynamical system
\begin{equation}\label{eq:nonhomogeneous_tensor_system}
    \dot{x} = -A x^{k-1} - B x^{m-1} - C x^{\ell-1} + \cdots.
\end{equation}
If each tensor $A\in \mathbb{R}^{[n,k]},B\in \mathbb{R}^{[n,m]},C\in \mathbb{R}^{[n,\ell]},\dots$ satisfies the condition of Lemma~\ref{lem:sufficient_positive}, then the system is positive.
\end{lemma}
\begin{proof}
The proof follows directly from the same analysis in Lemma \ref{lem:sufficient_positive}.
\end{proof}

We can generalize the positivity condition further by allowing a state-dependent diagonal scaling.

\begin{lemma}\label{lem:diag_f_positive}
Consider the tensor-based dynamical system
\begin{equation}\label{eq:diag_f_system}
\dot{x}=-\mathrm{diag}(f(x))Ax^{k-1},
\end{equation}
where $f(x)=(f_1(x_1),\dots,f_n(x_n))^\top$ is smooth and satisfies $f_i(0)=0$ for all $i\in[n]$. Then, for any tensor $A$, the nonnegative orthant $\mathbb{R}_+^n$ is forward invariant. Hence, the system \eqref{eq:diag_f_system} is positive.
\end{lemma}

\begin{proof}
For any $x\in\mathbb{R}_+^n$ such that $x_i=0$, we have $f_i(x_i)=f_i(0)=0$. Therefore,
$
    \dot{x}_i=-f_i(0)(Ax^{k-1})_i=0.
$
Thus, the vector field cannot point outside the nonnegative orthant through any of its boundary faces. Since the vector field is smooth, the nonnegative orthant $\mathbb{R}_+^n$ is forward invariant.
\end{proof}

\begin{remark}
A common choice is $f_i(x_i)=a x_i^b$ with $a>0$ and integer $b>0$. In this case,
$
    \mathrm{diag}(f(x))Ax^{k-1}
$
is a homogeneous polynomial vector field of degree $k+b-1$ and can therefore be represented as $\widetilde{A}x^{m-1}$ with $m=k+b$ for a suitable order-$m$ tensor $\widetilde{A}$. In particular, every monomial appearing in the $i$th component contains the factor $x_i^b$. Hence, the $i$th component vanishes whenever $x_i=0$, which guarantees positivity independently of the sign structure of the original tensor $A$.
\end{remark}

\subsection{Necessary and Sufficient Conditions}

The above lemmas provide convenient sufficient conditions. However, more general necessary and sufficient conditions can be derived based on the notion of copositivity.

\begin{definition}[(Strictly) copositive supersymmetric tensor]
Let $A\in\mathbb{R}^{[n,k]}$ be a supersymmetric tensor. 
We say that $A$ is \emph{copositive} if
    $A x^{k}\;\ge\;0,\qquad \forall\, x\in\mathbb{R}^n_+\setminus\{\mathbf{0}\},$
and \emph{strictly copositive} if
   $A x^{k}\;>\;0,\qquad \forall\, x\in\mathbb{R}^n_+\setminus\{\mathbf{0}\}.$
\end{definition}

Following \cite{wang2021hypergraph}, we introduce a symmetrization operator. For any $x=\left(x_1, \ldots, x_n\right)^{\top} \in \mathbb{R}^n$ satisfying $x_i=0$, we have
$
\left({A} x^{m-1}\right)_i=\sum_{i_2, \ldots, i_m \neq i} a_{i i_2 \cdots i_m} x_{i_2} \cdots x_{i_m} .
$
Let
$
\widetilde{x}=\left(x_1, \ldots, x_{i-1}, x_{i+1}, \ldots, x_n\right)^{\top} .
$
Then there exists a unique symmetric tensor $\operatorname{symm}\left(\mathscr{A}_i\right) \in \mathbb{R}^{[m-1, n-1]}$ such that
$
\operatorname{symm}\left(\mathscr{A}_i\right) \widetilde{x}^{m-1}=\left(\mathscr{A} x^{m-1}\right)_i=\sum_{i_2, \ldots, i_m \neq i} A_{i i_2 \cdots i_m} x_{i_2} \cdots x_{i_m}
$
that is, $\operatorname{symm}\left(\mathscr{A}_i\right) \widetilde{x}^{m-1}$ and $\left(\mathscr{A} x^{m-1}\right)_i$ are the same polynomial with respect to $\left\{x_k: k \neq i\right\}$.

\begin{lemma}[\!\cite{wang2021hypergraph}]\label{lem:copositive_condition}
For $k\ge3$, the system $\dot{x}=-A x^{k-1}$ is positive if and only if all $\mathrm{symm}(-A_i)$ are copositive.
\end{lemma}

This result provides a constructive criterion to verify positivity by checking the copositivity of symmetric tensors.

\begin{corollary}\label{cor:sum_positive}
If the system is composed of multiple homogeneous components as in \eqref{eq:nonhomogeneous_tensor_system}, and each tensor $A,B,C,\dots$ satisfies the copositivity condition in Lemma~\ref{lem:copositive_condition}, then the entire system is positive.
\end{corollary}

\begin{remark}
The condition in Lemma~\ref{lem:copositive_condition} is both necessary and sufficient, but generally difficult to verify. Lemma~\ref{lem:sufficient_positive} provides a simpler sufficient condition that can be easily checked from the sparsity pattern of the tensor.
\end{remark}

\section{Positive definite tensor}\label{sec:pdt}

The notion of a positive definite tensor originates from \cite{qit}.  
Let $A\in\mathbb{R}^{[n,m]}$ be a supersymmetric tensor of even order $m$.  
We call $A$ \emph{positive (semi-)definite} if
  $  A x^{m} > (\ge)\, 0, \quad \forall\, x \in \mathbb{R}^n \setminus \{\mathbf{0}\}.$
It has been shown in \cite{qit} that for an even-order supersymmetric tensor,  
both H-eigenvalues and Z-eigenvalues exist, and $A$ is positive (semi-)definite  
if and only if all of its H-eigenvalues (Z-eigenvalues) are positive (nonnegative).  It is obvious that an odd-order tensor cannot be positive (semi-)definite. 
If a homogeneous polynomial can be represented as a sum of squares (SOS),  
then the corresponding tensor is called an SOS tensor, which is necessarily positive semidefinite;  
however, the converse does not always hold.

For a supersymmetric tensor, one can check its H-eigenvalues through the Gershgorin circle theorem as follows.

\begin{lemma}[Theorem 6, \cite{qit}]\label{lem:circle}
The eigenvalues of a supersymmetric tensor $A$ lie in the union of $n$ disks in $\mathbb{C}$.  
Each disk is centered at a diagonal entry of $A$ and has as radius  
the sum of the absolute values of the corresponding off-diagonal entries.
\end{lemma}

\noindent A tensor $A$ is called \emph{diagonally dominant} (DD) if
\[
    |A_{ii\ldots i}|
    \ge \sum_{(i_2,\ldots,i_m)\neq(i,\ldots,i)} |A_{ii_2\ldots i_m}|,\quad
    i=1,\ldots,n;
\]
and \emph{strictly diagonally dominant} (SDD) if the inequalities above are strict.  
This leads to the following useful result.

\begin{lemma}\label{lem:check}
If $A$ is an even-order supersymmetric and strictly diagonally dominant tensor  
(diagonally dominant tensor) with all positive diagonal entries,  
then all H-eigenvalues of $A$ are positive, i.e., $A$ is positive (semi-)definite.
\end{lemma}


Positive definite tensors can be used in Lyapunov analysis of polynomial systems like the following.



\begin{thm}\label{thm:1}
Consider the tensor-based nonhomogeneous system
\[
    \dot{x}=-A x^{k-1}-B x^{m-1}-C x^{l-1}+\cdots,
\]
where $A,B,C,\ldots$ are positive semidefinite.  
Then the origin $\mathbf{0}$ is a stable equilibrium.  
If one of these tensors is positive definite,  
the origin is globally asymptotically stable.
\end{thm}

\begin{proof} 
Let $V(x)=\tfrac{1}{2}x^\top x>0$ for nonzero vector $x$. It is obvious that $V(x)$ is radially unbounded.  Then
    $\dot{V}=x^\top\dot{x}
    =x^\top(-Ax^{k-1}-Bx^{m-1}-Cx^{l-1}+\cdots)
    =-A x^{k}-B x^{m}-C x^{l}+\cdots \le 0.$
If at least one tensor is positive definite, then $\dot{V}<0$ for all $x\neq\mathbf{0}$,  
implying global asymptotic stability.
\end{proof}

A limitation of the definition in \cite{qit} is that positive definiteness requires symmetry.  
Following \cite{chen2013circulant}, this notion can be generalized to arbitrary even-order tensors via another type of symmetrization.

Let $A\in\mathbb{R}^{[n,m]}$.  
There exists a unique symmetric tensor $B\in\mathbb{R}^{[n,m]}$ such that  
$A x^{m}=B x^{m}$ for all $x\in\mathbb{R}^n$.  
We call $B$ the \emph{symmetrization} of $A$ and denote it by $\mathrm{sym}(A)$.  
When $m$ is even, $A$ is positive definite  
if and only if all H-eigenvalues of $\mathrm{sym}(A)$ are positive.

For any index set $(j_1,\ldots,j_m)$,  
let $\Sigma(j_1,\ldots,j_m)$ denote the set of all distinct permutations of the indices.  
Then $|\Sigma(j_1,\ldots,j_m)|$ equals $1$ if all indices are identical,  
and $m!$ if all indices are distinct.  
The entries of $B=\mathrm{sym}(A)$ are given by
\begin{equation}\label{eq:sym}
   B_{j_1\cdots j_m}
   =\frac{\sum_{\sigma\in\Sigma(j_1,\ldots,j_m)}A_{\sigma(j_1,\ldots,j_m)}}
          {|\Sigma(j_1,\ldots,j_m)|}.
\end{equation}
Although $A x^{m}=\mathrm{sym}(A)x^{m}$ for any $x$,  
it generally does not hold that $A x^{m-1}=\mathrm{sym}(A)x^{m-1}$.  
Hence the systems $\dot{x}=A x^{m-1}$ and $\dot{x}=\mathrm{sym}(A)x^{m-1}$  
are dynamically different.



\begin{thm}\label{thm:pdgo}
Consider the tensor-based nonhomogeneous system
\[
    \dot{x}=-A x^{k-1}-B x^{m-1}-C x^{l-1}+\cdots,
\]
where $\mathrm{sym}(A),\mathrm{sym}(B),\mathrm{sym}(C),\ldots$  
are positive semidefinite.  
Then the origin $\mathbf{0}$ is a stable equilibrium.  
If at least one of these symmetrized tensors is positive definite,  
the origin $\mathbf{0}$ is globally asymptotically stable.
\end{thm}

\begin{proof}The proof is similar to the Theorem \ref{thm:1} by using the same Lyapunov function.
\end{proof}

\section{Copositive Tensors}\label{sec:copositive_p_tensor}

In this section we investigate the (strictly) copositive tensors, 
and demonstrate how these concepts lead to stability guarantees for tensor-based systems 
evolving in the positive orthant. 
Throughout this section we retain the \emph{Positivity} Assumption \ref{ass:positivity}, 
meaning that trajectories starting in $\mathbb{R}^n_+$ remain within $\mathbb{R}^n_+$ for all time.

\subsection{(Strictly) copositive tensors}

The notion of copositivity extends the concept of positive definiteness to the nonnegative orthant. 
It plays an essential role when analyzing positive systems.

For a non-symmetric tensor $A$, we adopt the symmetrization-based extension consistent with Section~\ref{sec:pdt}: 
we say that $A$ is (strictly) copositive if and only if $\operatorname{sym}(A)$ is (strictly) copositive. 
This definition allows us to unify symmetric and non-symmetric cases under a single algebraic framework.

\subsection{Stability from (strict) copositivity in the positive orthant}

We now show that (strict) copositivity provides a simple Lyapunov-based condition 
for guaranteeing asymptotic stability of tensor-based polynomial systems restricted to $\mathbb{R}^n_+$. 
The next results parallel the positive definiteness results from Section~\ref{sec:pdt}, 
but are localized to the positive orthant.




\begin{prop}\label{prop:copositive_stability_nh}
Consider the tensor-based nonhomogeneous system
$\dot{x} \;=\; -A x^{k-1} \;-\; B x^{m-1} \;-\; C x^{\ell-1} \;+\; \cdots.$
Under the positivity assumption, if all of $A,B,C,\ldots$ are \emph{copositive} with one of those \emph{strictly copositive}, 
or all of $\operatorname{sym}(A),\operatorname{sym}(B),\operatorname{sym}(C),\ldots$ are \emph{copositive} with one of those \emph{strictly copositive}, 
then the origin $\mathbf{0}$ is \emph{globally asymptotically stable} in $\mathbb{R}^n_+$.
\end{prop}

\begin{proof}The proof is similar to the Theorem \ref{thm:1} by using the same Lyapunov function.
\end{proof}


\section{P-Tensor}\label{sec:ptensor}

In this section we introduce the notion of \emph{P-tensors}, which generalize the classical P-matrix concept to multilinear settings. 
This generalization enables the analysis of \emph{odd-order positive tensor-based systems}, 
where the usual notion of positive definiteness is not applicable. 
The P-tensor framework thus provides a unified tool for both even- and odd-order tensor systems.

\begin{definition}[P-tensor and $P_0$-tensor, {\cite{ding2015p}}]
A $k$-th order, $n$-dimensional tensor $A\in\mathbb{R}^{[n,k]}$ is called a \emph{P-tensor} if, for every nonzero $x\in\mathbb{R}^n$, there exists an index $i$ such that
$
    x_i^{\,k-1}\big(Ax^{k-1}\big)_i>0.
$
It is called a \emph{$P_0$-tensor} if, for every nonzero $x\in\mathbb{R}^n$, there exists an index $i$ such that
$
    x_i\neq0
    \quad\text{and}\quad
    x_i^{\,k-1}\big(Ax^{k-1}\big)_i\ge0.
$
When $k=2$, these definitions reduce to the standard notions of a P-matrix and a $P_0$-matrix, respectively.
\end{definition}

The above definition provides a way to characterize directional positivity in each component of a tensor mapping. 
This property allows P-tensors to serve as a bridge between algebraic positivity and dynamical stability.

\begin{lemma}[Properties of P-tensors {\cite{ding2015p}}]\label{lem:ptensor_properties}
Let $A\in\mathbb{R}^{[n,k]}$. The following statements hold:
\begin{enumerate}
    \item If $A$ is positive definite, then $A$ is a P-tensor.
    \item If $A$ is a nonsingular $H$-tensor with all positive diagonal entries, then $A$ is a P-tensor.
    \item If $A$ is a strictly diagonally dominant tensor with all positive diagonal entries, then $A$ is also a P-tensor.
\end{enumerate}
These algebraic conditions offer practical criteria for verifying the P-tensor property in applications.
\end{lemma}

\begin{assumption}[Positivity]\label{ass:positivity}
The dynamical system under consideration is \emph{positive}, 
meaning that the nonnegative orthant $\mathbb{R}^n_+$ is positively invariant. 
That is, trajectories starting in $\mathbb{R}^n_+$ remain within $\mathbb{R}^n_+$ for all $t\ge0$.
\end{assumption}

This assumption can be easily checked by using the conditions provided in Section \ref{sec:positive_tensor_system}.

\begin{lemma}\label{lem:P_is_strictly_copositive}
Any supersymmetric $P$-tensor is strictly copositive (see, e.g., \cite{ding2015p}).
\end{lemma}

The above lemma links two seemingly distinct concepts: 
$P$-tensors, defined via componentwise positivity conditions, 
and strictly copositive tensors, defined via positive multilinear forms on the nonnegative orthant. 
This connection naturally motivates the study of stability properties of systems governed by such tensors. Thus, we have the following corollary.

\begin{corollary}[Asymptotic stability in $\mathbb{R}^n_+$]\label{cor:ptensor_stability}
Consider the tensor-based multilinear system
\[
\dot{x} = -A x^{k-1},
\]
under Assumption~\ref{ass:positivity}, where $\operatorname{sym}(A)$ is a $P$-tensor.
Then the origin $\mathbf{0}$ is an \emph{asymptotically stable equilibrium} in the first orthant $\mathbb{R}^n_+$.
\end{corollary}

\begin{lemma}[H-eigenvalues of P- and $P_0$-tensors \cite{ding2015p}]
\label{lem:ptensor_heigenvalue}
A P-tensor has no non-positive H-eigenvalues, while a $P_0$-tensor has no negative H-eigenvalues.
\end{lemma}

Consequently, we have the following result.

\begin{corollary}[Global asymptotic stability for even $k$]\label{cor:ptensor_global_even}
Consider the same system with even order $k$,
 $\dot{x} = -A x^{k-1},$
where $\operatorname{sym}(A)$ is a P-tensor. 
Then the origin $\mathbf{0}$ is \emph{globally asymptotically stable} in $\mathbb{R}^n$.
\end{corollary}

\begin{proof}
By Lemma~\ref{lem:ptensor_heigenvalue}, all H-eigenvalues of the P-tensor $\operatorname{sym}(A)$ are positive. Since $k$ is even and $\operatorname{sym}(A)$ is supersymmetric, it is positive definite.
\end{proof}

\begin{remark}
The symmetrization condition in Corollary~\ref{cor:ptensor_global_even} is essential. In the linear matrix case, the stability of $\dot{x}=-Ax$ can be characterized directly by the real parts of the eigenvalues of $A$, since the solution is governed by the matrix exponential $e^{-At}$. For the higher-order tensor system $\dot{x}=-A x^{k-1}$, however, it generally has no analogous exponential solution and thus no analogous direct spectral characterization is available. With the quadratic Lyapunov function, one has
$
\dot V(x)=-A x^k=-\operatorname{sym}(A)x^k,
$
so the stability condition depends on the sign of a homogeneous polynomial. This polynomial is determined by the supersymmetric tensor $\operatorname{sym}(A)$, rather than by the real parts of the eigenvalues of the original tensor. 
\end{remark}

\section{Extended Tensor-eigenvalue Problem Characterization of Positivity}

This section establishes a novel spectral–variational characterization of the
positivity of general multivariate polynomials through the
\emph{extended Tensor-eigenvalue problem framework} (ETEP framework).
The formulation applies to nonhomogeneous polynomials and provides a
constructive sufficient condition for positive definiteness.

\subsection{Generalized Rayleigh formulation}

Consider a real polynomial
\[
P(x)=\sum_{j=m}^{k} A_j x^{j}, \qquad x\in\mathbb{R}^{n},
\]
where each $A_j\in\mathbb{R}^{[n,j]}$ is a supersymmetric tensor  according to Lemma \ref{lemma_1}.
Let $x=ru$ with $r=\|x\|_2>0$ and $u\in\mathbb{S}^{n-1}$.
Define the \emph{generalized Rayleigh map}
\begin{equation}\label{eq:lambda}
\lambda(u,r):=\frac{P(ru)}{r^{k}}
=\sum_{j=m}^{k}\frac{A_j u^{j}}{r^{\,k-j}},
\qquad (u,r)\in\mathbb{S}^{n-1}\times(0,\infty).
\end{equation}

The algebraic identity
\begin{equation}\label{eq:identity}
P(ru)=r^{\,k}\lambda(u,r)
\end{equation}
links the original polynomial to its scaled spectral representation.

\subsection{Extended Tensor Eigenvalue Problem (ETEP)}

Introduce the homogeneous lifting
\[
\widetilde P(x,t)=\sum_{j=m}^{k} t^{\,k-j} A_j x^{j},
\qquad (x,t)\in\mathbb{R}^{n}\times\mathbb{R}_{++}.
\]

The minimization of $\widetilde P$ on the unit sphere
$\{(x,t):\ \|x\|_2^2+t^2=1,\ t>0\}$ is governed by the Lagrangian
\[
\mathcal L(x,t,\lambda)
=\widetilde P(x,t)
-\frac{\lambda}{2}(\|x\|_2^2+t^2-1).
\]

The first-order optimality (KKT) conditions are
\begin{equation}\label{eq:EH}
\begin{split}
    &\nabla_x \widetilde P(x,t)=\lambda x, \\
&\partial_t \widetilde P(x,t)=\lambda t,
\\
&\|x\|_2^2+t^2=1,\ t>0.
\end{split}
\end{equation}

For each $j$, one has
\[
\nabla_x(A_jx^{j})=j\,A_jx^{j-1},
\]
and
\[
\partial_t(t^{\,k-j}A_jx^{j})=(k-j)t^{\,k-j-1}A_jx^{j}.
\]

Hence, the extended Tensor-eigen system is 
\begin{equation}\label{eq:EH-operator}
{
\begin{aligned}
&\sum_{j=m}^{k} j\,t^{\,k-j}A_jx^{j-1}=\lambda x,\\[4pt]
&\sum_{j=m}^{k} (k-j)\,t^{\,k-j-1}A_jx^{j}=\lambda t,\\[4pt]
&\|x\|_2^2+t^2=1,\quad t>0.
\end{aligned}}
\end{equation}

\paragraph{Relation to the original nonhomogeneous polynomial.}

The core idea of the Extended Tensor-Eigenvalue Problem (ETEP) framework is to analyze the positivity of general multivariate polynomials through a homogenization procedure. The following proposition establishes the fundamental relationship between the homogeneous and non-homogeneous polynomials.

\begin{prop}[Homogenization and Original Polynomial]\label{prop:2}
Let $P(x) = \sum_{j=m}^k A_j x^j$ be a real polynomial with $A_j \in \mathbb{R}^{[n,j]}$ being supersymmetric tensors, and let $\widetilde{P}(x,t) = \sum_{j=m}^k t^{k-j} A_j x^j$ be its homogeneous extension. Then:

\begin{enumerate}
    \item For any $x \neq 0$ and $r = \|x\| > 0$, $u = x/\|x\| \in \mathbb{S}^{n-1}$, the following identity holds:
    \begin{equation}
        P(x) = (r^2 + 1)^{k/2} \cdot \widetilde{P}(x', t'),
    \end{equation}
    where $(x', t') = \left(\dfrac{u}{\sqrt{1 + 1/r^2}}, \dfrac{1/r}{\sqrt{1 + 1/r^2}}\right)$ satisfies $\|(x',t')\| = 1$ and $t' > 0$.
    
    \item Conversely, for any $(x_0, t_0)$ with $\|(x_0, t_0)\| = 1$ and $0<t_0\neq 1$, there exists a unique $x \neq 0$ such that:
    \begin{equation}
        \widetilde{P}(x_0, t_0) = \frac{P(x)}{(r^2 + 1)^{k/2}},
    \end{equation}
    where $r = \|x\| = \frac{\|x_0\|}{t_0}$ and the corresponding $(x', t')$ equals $(x_0, t_0)$.
    
    \item The following sign-preserving bijection holds:
    \[
\operatorname{sgn} P(x) 
= 
\operatorname{sgn} \widetilde{P}(x',t'),
\qquad
\forall\, x\neq 0. 
\]

    \item The extended Tensor-eigen system \eqref{eq:EH-operator} captures all interior critical points of $\widetilde{P}$ on the hemisphere $\{(x,t) \mid \|(x,t)\| = 1, 0<t <1 \}$.
\end{enumerate}
\end{prop}

\begin{proof}
We prove each statement separately:

\textbf{Proof of (1):}
For any $x \neq 0$, let $r = \|x\| > 0$ and $u = x/r \in \mathbb{S}^{n-1}$. Define the normalized coordinates:
\[
(x', t') = \left(\frac{u}{\sqrt{1 + 1/r^2}}, \frac{1/r}{\sqrt{1 + 1/r^2}}\right).
\]
Direct computation shows $\|(x',t')\|^2 = \|x'\|^2 + (t')^2 = \frac{1}{1 + 1/r^2} + \frac{1/r^2}{1 + 1/r^2} = 1$, and clearly $t' > 0$.

Now, using the $k$-homogeneity ($k$ is the highest order) of $\widetilde{P}$:
$\widetilde{P}(x', t') = \widetilde{P}\left(\frac{u}{\sqrt{1 + 1/r^2}}, \frac{1/r}{\sqrt{1 + 1/r^2}}\right) 
= \frac{1}{(1 + 1/r^2)^{k/2}} \widetilde{P}(u, 1/r).$
From the definition of $\widetilde{P}$ and the relationship with $P$:
$\widetilde{P}(u, 1/r) = \sum_{j=m}^k (1/r)^{k-j} A_j u^j = \frac{1}{r^k} \sum_{j=m}^k r^j A_j u^j = \frac{P(ru)}{r^k} =\lambda(u,r)= \frac{P(x)}{r^k}.$

Therefore,
$\widetilde{P}(x', t') = \frac{1}{(1 + 1/r^2)^{k/2}} \cdot \frac{P(x)}{r^k} = \frac{P(x)}{r^k (1 + 1/r^2)^{k/2}} = \frac{P(x)}{(r^2 + 1)^{k/2}},$
which proves statement (1).

\textbf{Proof of (2):} 
For any $(x_0, t_0)$ with $\|(x_0, t_0)\| = 1$ and $0< t_0 \neq 1$ (If $t_0=1$, $x_0=0$ which contradicts  the assumption.), define:
\[
x = \frac{x_0}{t_0}, \quad r = \|x\| = \frac{\|x_0\|}{t_0} = \frac{\sqrt{1 - t_0^2}}{t_0}.
\]
Then $x \neq 0$ and $r > 0$. Let $u = x/r = x_0/\|x_0\|$.

Now construct $(x', t')$ as in (1):
\[
(x', t') = \left(\frac{u}{\sqrt{1 + 1/r^2}}, \frac{1/r}{\sqrt{1 + 1/r^2}}\right).
\]
Compute:
\[
1 + \frac{1}{r^2} = 1 + \frac{t_0^2}{1 - t_0^2} = \frac{1}{1 - t_0^2},
\]
so
\[
\sqrt{1 + \frac{1}{r^2}} = \frac{1}{\sqrt{1 - t_0^2}}.
\]
Then:
\[
x' = \frac{u}{\sqrt{1 + 1/r^2}} = u \cdot \sqrt{1 - t_0^2} = \frac{x_0}{\|x_0\|} \cdot \|x_0\| = x_0,
\]
\[
t' = \frac{1/r}{\sqrt{1 + 1/r^2}} = \frac{t_0/\sqrt{1 - t_0^2}}{1/\sqrt{1 - t_0^2}} = t_0.
\]
Thus $(x', t') = (x_0, t_0)$.

By statement (1), we have:
\[
P(x) = (r^2 + 1)^{k/2} \cdot \widetilde{P}(x', t') = (r^2 + 1)^{k/2} \cdot \widetilde{P}(x_0, t_0),
\]
which proves statement (2).

\textbf{Proof of (3):}
From statements (1) and (2), we have established a bijection between:
(a) $\mathbb{R}^n \setminus \{0\}$ (all nonzero $x$),
(b) $\{(x_0,t_0) : \|(x_0,t_0)\| = 1, 0<t<1\}$ (the open hemisphere without the north pole $(0,1)$)

Moreover, for any corresponding pair $(x, (x_0, t_0))$ under this bijection, we have:
\[
P(x) = (r^2 + 1)^{k/2} \cdot \widetilde{P}(x_0, t_0),
\]
where $(r^2 + 1)^{k/2} > 0$ is a positive scaling factor.

Therefore, the sign of $P(x)$ equals the sign of $\widetilde{P}(x_0, t_0)$ for all corresponding pairs, which proves the sign-preserving equivalences in statement (3).

\textbf{Proof of (4):}
The extended tensor-eigen system \eqref{eq:EH-operator} is derived from the first-order optimality conditions (KKT conditions) for minimizing $\widetilde{P}(x,t)$ subject to $\|(x,t)\|^2 = 1$ and $t > 0$. Specifically, the Lagrangian is:
\[
\mathcal{L}(x,t,\lambda) = \widetilde{P}(x,t) - \frac{\lambda}{2}(\|x\|_2^2 + t^2 - 1),
\]
and setting $\nabla_{(x,t)} \mathcal{L} = 0$ yields exactly equations \eqref{eq:EH-operator}. Therefore, solutions of \eqref{eq:EH-operator} correspond precisely to critical points of $\widetilde{P}$ on the specified hemisphere.
\end{proof}

\begin{remark}[Boundary cases]
The correspondence holds for interior points with $t>0$ (equivalently $r>0$).
Possible extrema at the boundaries $r\to0^+$ or $r\to\infty$
are excluded from this proposition and treated separately
through the limiting quantities $\alpha_\infty$ and $\beta_0$ introduced in the next subsection.
\end{remark}

%

\subsection{Definition of the interior spectral value}
We define the \emph{interior minimal spectral value} as
\begin{equation}\label{eq:lambda-int}
\lambda_{\mathrm{int}}
:=\inf\Big\{\lambda:\ ((x,t),\lambda)\ \text{satisfies}
\ \eqref{eq:EH-operator},\ \ x\neq 0\Big\},
\end{equation}
and adopt the convention that $\lambda_{\mathrm{int}}=+\infty$ if no such
solution exists.
Only the nontrivial cases $x\neq 0$ are considered, as the homogeneous
scaling in \eqref{eq:EH-operator} makes $(x,t)=(0,1)$ degenerate.

\subsection{Boundary regimes}

The global infimum of $\lambda(u,r)$ may occur at the boundary
$r\to0^+$ or $r\to\infty$.
Define the uniform boundary limits
\begin{equation}\label{eq:boundary-bounds}
\begin{split}
\alpha_\infty&:
=\inf_{\|u\|=1}A_k u^k, \\
\beta_0&:=\inf_{\|u\|=1}A_m u^m.
\end{split}
\end{equation}
The term $\alpha_\infty$ captures the asymptotic contribution of the leading
homogeneous part $A_kx^k$, while $\beta_0$ depends on the first nonzero
even-order term dominating as $r\to0^+$. These quantities should be interpreted as sign-determining boundary forms rather than as limits of the same normalized Rayleigh quotient.
If the lowest nonzero degree is odd, then $\beta_0\le0$. So is the highest degree, which must be even for positive (semi-)definiteness. All boundary cases can be easily checked by Z- or H-eigenproblem.

\subsection{Spectral criterion for positivity}

\begin{thm}[Extended Tensor-eigenvalue criterion]
\label{thm:PD}
Let $P(x)=\sum_{j=m}^{k}A_jx^j$ with
$A_j\in\mathbb{R}^{[n,j]}$, and define
\[
\underline{\lambda}:=\min\{\lambda_{\mathrm{int}},\,\alpha_\infty,\,\beta_0\}.
\]
Then:
\begin{enumerate}
\item If $\underline{\lambda}>0$, the polynomial $P$ is positive definite:
$P(x)>0$ for all $x\neq0$.
\item If $\underline{\lambda}\ge0$, the polynomial $P$ is positive
semidefinite:
$P(x)\ge0$ for all $x$.
Moreover, the PSD condition is equivalent:
$P(x)\ge0$ for all $x$ if and only if
$\underline{\lambda}\ge0$.
\end{enumerate}
The PD condition is sufficient but not necessary.
\end{thm}

\begin{proof}
By Proposition~\ref{prop:2}, for every $x\neq 0$, there exists a unique corresponding point $z=(x',t')$ with $\|z\|_2=1$ and $0<t'<1$ such that $P(x)=(1+\|x\|_2^2)^{k/2}\widetilde P(x',t')$. Since the scaling factor is strictly positive, $P(x)$ and $\widetilde P(x',t')$ always have the same sign.

Suppose first that $\underline{\lambda}\ge0$, i.e., $\lambda_{\mathrm{int}}\ge0$, $\alpha_\infty\ge0$, and $\beta_0\ge0$. Assume, by contradiction, that $P(x_0)<0$ for some $x_0\neq0$. By Proposition~\ref{prop:2}, the corresponding point $z_0$ in the open hemisphere satisfies $\widetilde P(z_0)<0$.

Consider $\widetilde P$ on the closed upper hemisphere $\mathcal H=\{(x,t):\|x\|_2^2+t^2=1,\ t\ge0\}$. Since $\mathcal H$ is compact, $\widetilde P$ attains its minimum on $\mathcal H$. On the equator $t=0$, one has $\widetilde P(x,0)=A_kx^k\ge\alpha_\infty\ge0$, while at the north pole $(0,1)$ one has $\widetilde P(0,1)=0$. Hence, because $\widetilde P(z_0)<0$, a negative minimum must be attained at an interior point $z^\star=(x^\star,t^\star)$ with $0<t^\star<1$ and $x^\star\neq0$.

Such an interior minimizer satisfies the KKT conditions in~(10). Since $\widetilde P$ is homogeneous of degree $k$, Euler's identity gives $(z^\star)^\top\nabla\widetilde P(z^\star)=k\widetilde P(z^\star)$. Combining this with $\nabla\widetilde P(z^\star)=\lambda^\star z^\star$ and $\|z^\star\|_2=1$ yields $\lambda^\star=k\widetilde P(z^\star)<0$. This contradicts $\lambda_{\mathrm{int}}\ge0$. Therefore, $P(x)\ge0$ for all $x$.

Conversely, suppose that $P(x)\ge0$ for all $x$. By Proposition~\ref{prop:2}, $\widetilde P(z)\ge0$ at every corresponding interior point. Hence, for every interior solution of~(10), Euler's identity gives a nonnegative eigenvalue parameter, and therefore $\lambda_{\mathrm{int}}\ge0$.

Moreover, if $\alpha_\infty<0$, then there exists $u\in\mathbb S^{n-1}$ such that $A_ku^k<0$. Since $A_ku^k$ is the leading-order term of $P(ru)$, one would have $P(ru)<0$ for sufficiently large $r$, which is a contradiction. Thus, $\alpha_\infty\ge0$. Similarly, if $\beta_0<0$, then there exists $u\in\mathbb S^{n-1}$ such that $A_mu^m<0$. Since $A_mu^m$ is the lowest-order nonzero term of $P(ru)$, one would have $P(ru)<0$ for sufficiently small $r>0$, again giving a contradiction. Hence, $\beta_0\ge0$. Therefore, $\underline{\lambda}\ge0$.

Finally, suppose that $\underline{\lambda}>0$. From the preceding result, $P$ is positive semidefinite. Assume that $P(x_0)=0$ for some $x_0\neq0$. Proposition~2 then gives a corresponding interior point $z_0$ such that $\widetilde P(z_0)=0$. Since $\widetilde P$ is nonnegative throughout the open hemisphere, $z_0$ is an interior minimizer and therefore satisfies~(10). Euler's identity then gives the corresponding eigenvalue parameter $\lambda_0=0$, contradicting $\lambda_{\mathrm{int}}>0$. Hence, $P(x)>0$ for all $x\neq0$, and $P$ is positive definite.
\end{proof}




\begin{remark}
The extended Tensor-eigen equations \eqref{eq:EH-operator} capture only
\emph{interior} stationary points where the gradient vanishes.
When the minimum of $\lambda(u,r)$ occurs at
$r\to0^+$ or $r\to\infty$, no finite solution of
\eqref{eq:EH-operator} exists.
Theorem~\ref{thm:PD} therefore combines the interior spectral value
$\lambda_{\mathrm{int}}$ with the boundary quantities
$\alpha_\infty$ and $\beta_0$ to provide a complete sufficient
criterion for positivity. For checking the boundary quantities, as they are governed by a homogeneous polynomial, we suggest using the classical tensor H-(Z-)eigenvalues to check the boundary quantities. Similar to the computations of H-(Z-)eigenvalues, our proposed procedure is also NP-hard.
\end{remark}

\begin{remark}
The extended tensor–eigenvalue test is conservative:
it may occasionally classify a strictly positive polynomial as only
semidefinite when the smallest spectral value $\underline{\lambda}$
is numerically close to zero, for example when $P(x)$ is nearly flat
along some direction.
Such cases correspond to weak curvature or boundary degeneracy in the
Rayleigh map, where the true minimum is positive but attained in the
limit $r\!\to\!0$ or $r\!\to\!\infty$.
Nevertheless, the method never yields false positives, it may miss PD as PSD
but never misclassify a non-PSD polynomial as PSD.
\end{remark}

\subsection{Equivalent Z-eigenproblem formulation}
\label{subsec:z_equiv}

The extended system~\eqref{eq:EH-operator} admits an equivalent representation
as a standard Z-eigenvalue problem of a higher-dimensional supersymmetric tensor.
This reformulation clarifies that the \emph{interior} eigenpairs $((x,t),\lambda)$
(i.e., those with $0<t<1$) are precisely the Z-eigenpairs of a homogenized tensor
constructed from $\{A_j\}$, restricted to the open hemisphere.

\paragraph{Construction of the extended tensor.}
Let $z:=(x,t)\in\mathbb{R}^{n+1}$ and
denote by $e_t : = (0,\ldots,0,1)^\top \in \mathbb{R}^{n+1}$ the unit vector corresponding to the
last coordinate. Define the order-$k$ supersymmetric tensor
\begin{equation}\label{eq:extended_tensor}
\mathcal{T}
=\sum_{j=m}^{k}\frac{1}{\binom{k}{j}}\,
\mathrm{sym}\!\big(A_j\otimes e_t^{\otimes (k-j)}\big).
\end{equation}
Since each $A_j$ is supersymmetric, $\mathcal{T}$ is also supersymmetric and satisfies
\begin{equation}\label{eq:Te_hom_identity}
\widetilde P(x,t)
=\mathcal{T}z^{k}.
\end{equation}

\paragraph{Z-eigenvalue equivalence (interior region).}
Differentiating~\eqref{eq:Te_hom_identity} and comparing with~\eqref{eq:EH-operator} yields that
a pair $((x,t),\lambda)$ with $\|z\|_2=1$ and $0<t<1$ solves~\eqref{eq:EH-operator}
if and only if it satisfies the Z-eigenvalue equations
\begin{equation}\label{eq:z_equiv}
\mathcal{T}z^{k-1}=\lambda z,\qquad
\|z\|_2=1,\quad 0<t<1.
\end{equation}
Hence, each interior solution $((x,t),\lambda)$ of~\eqref{eq:EH-operator} corresponds exactly to a
Z-eigenpair $(z,\lambda)$ of $\mathcal{T}$ with $z=(x,t)$ and $0<t<1$.
Conversely, any Z-eigenpair $(z,\lambda)$ of $\mathcal{T}$ whose eigenvector satisfies $0<t<1$
satisfies~\eqref{eq:EH-operator}.
Therefore, the interior spectral value $\lambda_{\mathrm{int}}$ defined in~\eqref{eq:lambda-int}
coincides with the smallest Z-eigenvalue of $\mathcal{T}$ \emph{over the interior set}:
\begin{equation}\label{eq:lambda_int_as_restricted_Z}
\lambda_{\mathrm{int}}
=\inf\Big\{\lambda:\ \mathcal{T}z^{k-1}=\lambda z,\ \|z\|_2=1,\ 0<t<1\Big\}.
\end{equation}

\begin{remark}
    Any scalar factor $k$ arising from differentiating the degree-$k$ homogeneous polynomial can be absorbed into the eigenvalue parameter by a rescaling of $\lambda$; hence, it does not affect the corresponding eigenvectors or the above spectral equivalence. Here, $\lambda$ is understood to denote the rescaled eigenvalue under this convention.
\end{remark}

\begin{remark}[Boundary geometry: north pole, equator]
\label{rem:north_pole}
The restriction $0<t<1$ (equivalently, $t>0$ and $x\neq 0$ on $\|z\|_2=1$) is essential.
The ``north pole'' $z=e_t=(0,\ldots,0,1)$ corresponds to $x=0$ under homogenization, and it is a
degenerate boundary point excluded from the interior definition of $\lambda_{\mathrm{int}}$.
In particular, when $P(0)=0$ (no constant term), one has $\widetilde P(0,1)=0$ and typically
$\mathcal{T}e_t^{k-1}=0$, so $(e_t,\lambda=0)$ becomes a Z-eigenpair of $\mathcal{T}$.
Including this point would force the \emph{unrestricted} smallest Z-eigenvalue of $\mathcal{T}$ to be
$\le 0$, even when $\widetilde P(x,t)>0$ for all $t>0$ and $x\neq 0$.
Besides the north pole, the equator
$\{(x,t): \|z\|_2=1,\ t=0\}$
forms another boundary component of the closed hemisphere, corresponding to the asymptotic regime
$r\to\infty$. Its contribution is not captured by the restricted interior problem either, and is treated
separately through the boundary quantity $\alpha_\infty$.
Likewise, the north-pole limit $t\to 1$ (equivalently, $r\to 0^+$) is accounted for by $\beta_0$. Here, $(e_t,0)$ mentioned above should be understood as a degenerate artificial root induced by the homogenization at the north pole, rather than a meaningful boundary minimizer; by contrast, the relevant boundary contribution near the north pole arises from the nontrivial limit $t\to 1$, which is quantified by $\beta_0$.
For this reason, $\lambda_{\mathrm{int}}$ is a \emph{restricted} interior quantity as in~\eqref{eq:lambda_int_as_restricted_Z},
and the global certificate in Theorem~\ref{thm:PD} must additionally account for the boundary regimes.
\end{remark}


\begin{remark}[Unified Z-eigenproblem perspective]
The positivity test in Theorem~\ref{thm:PD} is implemented through three Z-eigenvalue subproblems:
\begin{enumerate}
    \item the restricted interior Z-eigenproblem of the extended tensor $\mathcal T$ on
    $0<t<1$, which yields $\lambda_{\mathrm{int}}$;
    \item the Z-eigenproblem of the leading homogeneous tensor $A_k$, which yields
    $\alpha_\infty$ and corresponds to the equatorial boundary regime $t=0$
    (equivalently, $r\to\infty$);
    \item the Z-eigenproblem of the lowest nonzero even-order tensor $A_{j_0}$, which yields
    $\beta_0$ and corresponds to the north-pole limit $t\to 1$
    (equivalently, $r\to 0^+$).
\end{enumerate}
Accordingly, the final certificate is
$\underline{\lambda}=\min\{\lambda_{\mathrm{int}},\alpha_\infty,\beta_0\}.$
Conceptually, however, these three quantities are not independent: they can be viewed as arising from a
single spectral minimization problem for the homogenized polynomial on the closed hemisphere
$\{z=(x,t):\|z\|_2=1,\ t\ge 0\},$
where the minimum may be attained either in the interior $0<t<1$, on the equator $t=0$, or in the
north-pole limit $t\to 1$. It therefore can be viewed as one single Z-eigenproblem of the extended tensor $\mathcal T$.
In practical computation, nevertheless, it is preferable to evaluate the three subproblems separately.
Indeed, although the unified viewpoint is theoretically natural, directly solving the restricted interior
eigenproblem of the extended tensor may become numerically sensitive when the tensor order or dimension is
large; by contrast, the boundary quantities $\alpha_\infty$ and $\beta_0$ reduce to standard Z-eigenvalue
problems for homogeneous tensors and are usually more robust to compute separately with existing solvers
such as TenEig \cite{teneig}.
\end{remark}

\subsection{Stability of nonhomogeneous polynomial vector fields via ETEP}
\label{subsec:stability_via_etep}

Consider a polynomial vector field
\begin{equation}\label{eq:polyVF}
    \dot x \;=\; f(x)\;=\;\sum_{j=1}^{k} B_j x^{\,j-1},\qquad
    B_j\in\mathbb{R}^{[n,j]},
\end{equation}
where each $B_j$ is almost-symmetric in the last $j\!-\!1$ modes. Take the quadratic Lyapunov function
$V(x)=\tfrac12\|x\|_2^2$. Then
\begin{equation}\label{eq:Vdot_raw}
    \dot V(x)\;=\;x^\top f(x)
    \;=\;\sum_{j=1}^{k} x^\top B_j x^{\,j-1}.
\end{equation}

\paragraph{Symmetric scalarization.}
For each $j$, there exists a \emph{unique} supersymmetric tensor
$A_j\in\mathbb{R}^{[n,j]}$ such that
\begin{equation}\label{eq:scalarization}
    x^\top B_j x^{\,j-1}\;=\;A_j x^{\,j}\quad \text{for all }x, A_j=\mathrm{sym}(B_j).
\end{equation}

Consequently,
\begin{equation}\label{eq:Vdot_poly}
    \dot V(x)\;=\;\sum_{j=1}^{k} A_j x^{\,j}\,,
    \qquad
    S(x)\;:=\;-\dot V(x)\;=\;-\sum_{j=1}^{k} A_j x^{\,j}.
\end{equation}

\paragraph{ETEP-based stability test.}
Apply the extended tensor–eigenvalue framework to the scalar polynomial
$S(x)$.
Let $\lambda_{\mathrm{int}}(S)$, $\alpha_\infty(S)$, and $\beta_0(S)$ be
the three spectral quantities defined in \eqref{eq:lambda-int} and
\eqref{eq:boundary-bounds} for $S$. Define
\[
\underline{\lambda}(S)\;:=\;\min\{\lambda_{\mathrm{int}}(S),\,\alpha_\infty(S),\,\beta_0(S)\}.
\]
Then:
{\small\begin{equation}\label{eq:stability_criterion}
\boxed{
\begin{aligned}
\underline{\lambda}(S) \;>\; 0
&\;\Longrightarrow\;
\dot V(x)\;< 0
\ \Rightarrow\ \text{global asymptotic stability},\\[2pt]
\underline{\lambda}(S) \;\ge\; 0
&\;\Longrightarrow\;
\dot V(x)\;\le\;0
\ \Rightarrow\ \text{Lyapunov stability}.\\[2pt]
\end{aligned}}
\end{equation}}


\section{U-eigenvalues, unfolding representations, and their relations with Z- and H-eigenvalues}

In the previous sections, Z-eigenvalues and H-eigenvalues have been used to characterize
the positivity of homogeneous tensor forms and the boundary quantities arising in the ETEP
framework. To motivate the unfolding-based route adopted later, we now introduce
U-eigenvalues for even-order tensors through matrix unfoldings, and clarify their relations
with the classical Z- and H-eigenvalues. Here, we mainly focus on even-order tensors, since an odd-order tensor cannot be positive (semi-)definite.

\subsection{Even-order tensors and $(p,p)$-unfolding}

Let $m=2p$ be even and let $A\in \mathbb{R}^{[n,2p]}$.
For a vector $x\in \mathbb{R}^n$, the $m$-fold tensor product of $x$ is denoted by
$x^{\otimes m}:=\underbrace{x\otimes x\otimes\cdots\otimes x}_{m\ \text{times}}.$
In coordinates, the entries of $x^{\otimes 2p}$ are
$(x^{\otimes 2p})_{i_1\cdots i_{2p}}=x_{i_1}\cdots x_{i_{2p}}.$
Accordingly, the homogeneous polynomial associated with $A$ is written as
$Ax^{2p}
=
\sum_{i_1,\dots,i_{2p}=1}^n a_{i_1\cdots i_{2p}}x_{i_1}\cdots x_{i_{2p}}.$

Fix a bijection between multi-indices $(i_1,\dots,i_p)\in [n]^p$ and scalar indices in $[n^p]$.
The corresponding $(p,p)$-unfolding of $A$ is the matrix
$A^{(p)}\in \mathbb{R}^{n^p\times n^p}$ defined by
\[
(A^{(p)})_{\alpha,\beta}
=
a_{i_1\cdots i_pj_1\cdots j_p},
\]
whenever $\alpha \leftrightarrow (i_1,\dots,i_p)$ and
$\beta \leftrightarrow (j_1,\dots,j_p)$.

Now define the lifted vector $v(x)\in \mathbb{R}^{n^p}$ by
\[
v_\alpha(x)=x_{i_1}\cdots x_{i_p},
\qquad
\alpha\leftrightarrow (i_1,\dots,i_p),
\]
or equivalently,
\[
v(x)=\mathrm{vec}(x^{\otimes p}).
\]
Then the homogeneous polynomial $Ax^{2p}$ admits the quadratic-form representation
\begin{equation}
Ax^{2p}=v(x)^\top A^{(p)}v(x), \qquad \forall x\in \mathbb{R}^n .
\label{eq:unfolding_quadratic_form}
\end{equation}

Moreover, if $\|x\|_2=1$, then
\begin{equation}
\|v(x)\|_2^2
=
\sum_{i_1,\dots,i_p=1}^n (x_{i_1}\cdots x_{i_p})^2
=
\left(\sum_{i=1}^n x_i^2\right)^p
=
\|x\|_2^{2p}
=
1.
\label{eq:v_norm_identity}
\end{equation}

\begin{remark}[Symmetry of the $(p,p)$-unfolding matrix]
\label{rem:unfolding_symmetric}
If $A\in\mathbb{R}^{[n,2p]}$ is supersymmetric, then any $(p,p)$-unfolding $A^{(p)}$ is a symmetric matrix. Indeed, for any two multi-indices
$I=(i_1,\dots,i_p)$ and $J=(j_1,\dots,j_p)$, one has
$(A^{(p)})_{\phi(I),\phi(J)}
=
a_{i_1\cdots i_pj_1\cdots j_p}
=
a_{j_1\cdots j_pi_1\cdots i_p}
=
(A^{(p)})_{\phi(J),\phi(I)},$
where the middle equality follows from the supersymmetry of $A$.
\end{remark}

\subsection{Definition of U-eigenvalues}

The quadratic-form representation \eqref{eq:unfolding_quadratic_form} naturally leads to
a matrix-based eigenvalue notion for even-order tensors.

\begin{definition}[U-eigenvalues]
Let $A\in \mathbb{R}^{[n,2p]}$ and let $A^{(p)}$ be one of its $(p,p)$-unfoldings.
A scalar $\lambda_U\in \mathbb{R}$ is called a U-eigenvalue of $A$ if there exists a nonzero
vector $y\in \mathbb{R}^{n^p}$ such that
$A^{(p)}y=\lambda_U y.$
Any corresponding nonzero vector $y$ is called a U-eigenvector.
\end{definition}

Thus, U-eigenvalues are simply the matrix eigenvalues of the unfolding $A^{(p)}$.
In particular, when $A^{(p)}$ is symmetric, all U-eigenvalues are real and can be ordered.

\subsection{Invariance with respect to the choice of unfolding}

Since the $(p,p)$-unfolding depends on the ordering of the multi-indices,
one must verify that the above definition is independent of the particular unfolding chosen.
This is indeed the case for supersymmetric even-order tensors.

\begin{thm}[Invariance of U-eigenvalues]
\label{thm:U_invariance}
Let $A\in \mathbb{R}^{[n,2p]}$ be supersymmetric.
Then, the U-eigenvalues of a supersymmetric even-order tensor are well defined, i.e.,
independent of the particular choice of the $(p,p)$-unfolding.
\end{thm}

\begin{proof}
Let $M_{\phi}$ and $M_{\psi}$ be two $(p,p)$-unfoldings associated with two bijections
$\phi,\psi:[n]^p\to [n^p]$.
Since $\phi$ and $\psi$ are bijections on the same finite set, there exists a permutation
$\tau$ of $[n^p]$ such that $\psi=\tau\circ \phi$.
Let $P$ be the permutation matrix associated with $\tau$.

For any two $p$-tuples $I=(i_1,\dots,i_p)$ and $J=(j_1,\dots,j_p)$, let
$\alpha=\phi(I)$ and $\beta=\phi(J)$.
Then $\psi(I)=\tau(\alpha)$ and $\psi(J)=\tau(\beta)$, and by the definition of the unfolding,
$(M_{\psi})_{\tau(\alpha),\tau(\beta)}
=
a_{i_1\cdots i_pj_1\cdots j_p}
=
(M_{\phi})_{\alpha,\beta}.$
Hence
$M_{\psi}=P^\top M_{\phi}P.$
Therefore $M_{\phi}$ and $M_{\psi}$ are permutation similar and so have identical eigenvalues.
\end{proof}

\begin{remark}
The above argument also shows that, for supersymmetric tensors, the spectrum of the unfolding
is invariant under permutations of tensor modes before the $(p,p)$ split.
Thus the U-eigenvalues are intrinsic spectral quantities of the tensor, not artifacts of a
particular matricization convention.
\end{remark}

\subsection{Relations among U-, and Z-eigenvalues}

We now establish the relations between U-eigenvalues and the Z-/H-eigenvalues introduced earlier.

Recall that for a real symmetric even-order tensor $A\in \mathbb{R}^{[n,2p]}$,
the largest Z-eigenvalue admits the Rayleigh characterization
\begin{equation}
\lambda_Z^{\max}(A)
=
\max_{\|x\|_2=1} Ax^{2p}.
\label{eq:lambdaZmax_rayleigh}
\end{equation}

The next theorem shows that the U-eigenvalues of the unfolding provide natural upper bounds
for the Z-eigenvalues.

\begin{thm}[Relations among U-, and Z-eigenvalues]
\label{thm:UZH_relation}
Let $A\in \mathbb{R}^{[n,2p]}$ be supersymmetric and let $A^{(p)}$ be a $(p,p)$-unfolding matrix.
Then, 
    \begin{equation}
    \lambda_Z^{\max}(A)\le \lambda_{\max}(A^{(p)})=: \lambda_U^{\max}(A).
    \label{eq:Z_le_U}
    \end{equation}
\end{thm}

\begin{proof}
Let $\|x\|_2=1$. By \eqref{eq:unfolding_quadratic_form} and \eqref{eq:v_norm_identity},
\[
Ax^{2p}=v(x)^\top A^{(p)}v(x), \qquad \|v(x)\|_2=1.
\]
Therefore
$\lambda_Z^{\max}(A)
=
\max_{\|x\|_2=1} v(x)^\top A^{(p)}v(x)
\le
\max_{\|y\|_2=1} y^\top A^{(p)}y
=
\lambda_{\max}(A^{(p)}),$
which proves \eqref{eq:Z_le_U}.
\end{proof}

\begin{remark}[Eigenvectors generally do not coincide]
\label{rem:eigvec_not_same}
Although Theorem~\ref{thm:UZH_relation} compares the extremal values of U-, Z-, and H-eigenvalues,
their eigenvectors live in different spaces.
Indeed, a U-eigenvector is a vector in the lifted space $\mathbb{R}^{n^p}$,
whereas a Z-eigenvector belongs to $\mathbb{R}^n$.
More precisely, the Z-eigenvalue optimization only probes the quadratic form
$y^\top A^{(p)}y$ on the structured subset
\[
\{v(x)=\mathrm{vec}(x^{\otimes p}) : \|x\|_2=1\}\subset \mathbb{R}^{n^p},
\]
which is typically a very thin subset of the full unit sphere in $\mathbb{R}^{n^p}$.
Therefore, a U-eigenvector corresponds to a Z- or H-eigenvector only in exceptional cases where
it happens to be representable as $v(x)$ for some $x\in \mathbb{R}^n$.
\end{remark}

A similar result of a minimum-version is presented in the following.
\begin{corollary}[Minimum-version]
\label{cor:UZH_relation_min}
Let $A\in \mathbb{R}^{[n,2p]}$ be supersymmetric and let $A^{(p)}$ be a $(p,p)$-unfolding.
Then, $\lambda_Z^{\min}(A)\ge \lambda_{\min}(A^{(p)})=: \lambda_U^{\min}(A).$.
\end{corollary}

\begin{proof}
The result follows immediately by applying Theorem~\ref{thm:UZH_relation} to the tensor $-A$.
\end{proof}

\subsection{Implications for the ETEP framework}

The above bridge is particularly useful for the ETEP-based analysis and controller synthesis
developed later.
In the ETEP framework, positivity is characterized through restricted interior and boundary
Z-eigenvalue problems.
However, these Z-eigenvalue computations are generally difficult, especially for high-order lifted tensors.
The unfolding-based U-eigenvalue route replaces them by matrix spectral conditions on the
$(p,p)$-unfolding, thereby providing tractable sufficient certificates.
More specifically, for an even-order homogeneous tensor form $Ax^{2p}$,
Theorem~\ref{thm:UZH_relation} shows that nonnegativity of the unfolding matrix yields a sufficient condition for Z-eigenvalue nonnegativity.
At the same time, Remark~\ref{rem:eigvec_not_same} explains why such matrix-based surrogates
are generally conservative: they replace a structured tensor spectral problem by a relaxation
over the whole lifted Euclidean space.

\section{ETEP-guided U-eigenvalue controller synthesis}
\label{sec:ETEP_Ueig}

In the ETEP framework of Section VII, positivity of the polynomial
$S(x):=-\dot V(x)$
is characterized through three spectral quantities:
the interior quantity $\lambda_{\mathrm{int}}$, the high-order boundary quantity $\alpha_\infty$,
and the low-order boundary quantity $\beta_0$.
These quantities are defined through restricted Z-eigenvalue problems of the extended tensor
and of the boundary tensors.
However, solving such tensor eigenvalue problems exactly is in general computationally difficult.
This motivates us to introduce a tractable synthesis route based on U-eigenvalue/unfolding relaxations.

The key point is that, in the Lyapunov-derivative setting considered here, the homogenized polynomial
$S_e(x,t)=\sum_{j=m}^k t^{k-j}A_jx^j$
has no constant term whenever $S(0)=0$.
Hence the north pole $z=e_t=(0,\dots,0,1)$ always satisfies
$S_e(0,1)=0.$
As already discussed in Remark~7, this point is excluded from the restricted interior Z-eigenvalue
definition of $\lambda_{\mathrm{int}}$, but it remains present in any unrestricted unfolding representation
of the extended tensor. In  practice, available toolboxes are based on unrestricted settings. 
Therefore, for the extended tensor itself, one cannot in general expect a useful \emph{strictly positive}
unfolded spectral margin.
The natural role of the U-eigenvalue route is therefore to provide near \emph{PSD} certificates for the lifted
or higher-order part, while strict positivity is recovered separately through a low-order PD anchor.

\subsection{Unfolding of the extended tensor}

Assume that the highest degree $k$ is even, say $k=2p$.
Let
$T\in \mathbb{R}^{[n+1,k]}$
be the supersymmetric extended tensor defined in Section VII such that
$S_e(x,t)=Tz^k,\qquad z:=(x,t)\in \mathbb{R}^{n+1}.$
Fix a $(p,p)$-unfolding of $T$, denoted by $T^{(p)}\in \mathbb{R}^{(n+1)^p\times (n+1)^p}$.
Since $T$ is supersymmetric, $T^{(p)}$ is symmetric.
Define
$v(z):=\mathrm{vec}(z^{\otimes p}).$
Then one has
\begin{equation}
Tz^k = v(z)^\top T^{(p)} v(z),
\label{eq:extended_unfolding_identity}
\end{equation}
and, moreover,
\begin{equation}
\|v(z)\|_2^2=\|z\|_2^{2p}.
\label{eq:vz_norm}
\end{equation}

Therefore, if
\begin{equation}
T^{(p)} \succeq 0,
\label{eq:Tp_psd}
\end{equation}
then
\[
S_e(x,t)=Tz^k=v(z)^\top T^{(p)}v(z)\ge 0,\qquad \forall z\in \mathbb{R}^{n+1}.
\]
In particular, $S_e(x,t)\ge 0$ for all $(x,t)$ with $t>0$, and hence $S(x)\ge 0$ for all $x$.

\subsection{On the unfolding condition}

Although \eqref{eq:Tp_psd} is a valid sufficient condition, it can be conservative. Recall that $S_e(z)=v(z)^\top T^{(p)}v(z)$, where $v(z)=\operatorname{vec}(z^{\otimes p})$. The positivity of $S_e$ only requires this quadratic form to be nonnegative for realizable lifted vectors $v(z)$. In other words, the corresponding optimization is restricted to the structured set $\{v(z)=\operatorname{vec}(z^{\otimes p}):\|z\|_2=1\}$, where the representation $v(z)=\operatorname{vec}(z^{\otimes p})$ constrains the admissible directions in the lifted space.

By contrast, the minimum U-eigenvalue of $T$ is obtained from the unrestricted matrix problem over all unit vectors in the lifted space. More precisely,
\[
\min_{\|z\|_2=1} v(z)^\top T^{(p)}v(z)
\;\ge\;
\min_{\|y\|_2=1} y^\top T^{(p)}y
=
\lambda_{\min}(T^{(p)}).
\]
The vector $y$ in the U-eigenvalue problem is not required to satisfy $y=v(z)$ for any $z$. Therefore, even if the original polynomial, and consequently its homogeneous lifting, is PSD, $T^{(p)}$ may still have negative eigenvalues corresponding to directions that are not realizable by $v(z)$. Hence, $T^{(p)}\succeq0$ is a sufficient but in general not necessary condition for the nonnegativity of the polynomial.

There is also an additional restriction introduced by homogenization. Since $S(0)=0$, we have $S_e(0,1)=0$. At the north pole $z=e_t$, this gives $v(e_t)^\top T^{(p)}v(e_t)=0$, and hence $\lambda_{\min}(T^{(p)})\leq0$. Therefore, a strictly positive-definite unfolding $T^{(p)}\succeq\gamma I$ with $\gamma>0$ is impossible. Even in the best case, its smallest eigenvalue must be zero.

Thus, the conservatism comes from two sources. First, the unfolding-based U-eigenvalue problem replaces the structured optimization over $v(z)=\operatorname{vec}(z^{\otimes p})$ by an unrestricted optimization over the whole lifted space. Second, homogenization introduces additional degenerate directions, such as the north pole, which do not correspond to nonzero states of the original polynomial. These restrictions arise from the lifting and unfolding procedures rather than from the positivity of the original polynomial itself.

To reduce the effect of this conservatism, in the next subsection we split $S(x)$ into a positive-definite component and a remaining component whose unfolding is only required to be near-PSD. The former provides a strict local stabilizing margin, while the latter allows us to retain the computationally efficient unfolding-based SDP relaxation without requiring exact PSD of the extended unfolding.

\subsection{Split design: PD anchor plus PSD unfolded part}
To cope with the issue brought by the homogenization, we propose the following split design.
Let $j_0$ be the smallest nonzero even degree appearing in $S(x)$.
We decompose
\begin{equation}
S(x)=S_{\mathrm{pd}}(x)+S_{\mathrm{psd}}(x),
\label{eq:S_split_ueig}
\end{equation}
where
\begin{equation}
S_{\mathrm{pd}}(x):=\alpha A_{j_0}x^{j_0};
S_{\mathrm{psd}}(x):=(1-\alpha) A_{j_0}x^{j_0}+\sum_{\substack{j=j_0+1}}^k A_jx^j .
\label{eq:split_pd_psd}
\end{equation}
Here $S_{\mathrm{pd}}$ is chosen as the low-order PD anchor, while $S_{\mathrm{psd}}$ is treated by
unfolding/U-eigenvalue-based near PSD terms.

The rationale is simple:
\begin{itemize}
    \item the anchor term $A_{j_0}x^{j_0}$ is constrained to be positive definite, ensuring local stability;
    \item the higher-order part only needs to be near nonnegative to ensure a sufficiently big domain of attraction.
\end{itemize}

\subsection{U-eigenvalue constraints for the near-PSD part}

We now apply unfolding-based near-PSD constraints to the remaining part. Let $S_{e,\mathrm{psd}}(x,t)$ denote the homogenization of $S_{\mathrm{psd}}(x)$ to degree $k$, and let $T_{\mathrm{psd}}\in\mathbb{R}^{[n+1,k]}$ be the corresponding supersymmetric extended tensor, so that $S_{e,\mathrm{psd}}(x,t)=T_{\mathrm{psd}}z^k$. Let $M_{\mathrm{int}}$ be a $(p,p)$-unfolding of $T_{\mathrm{psd}}$.

A separate unfolding constraint for the leading-order tensor $A_k$ is not required here. Indeed, its $(p,p)$-unfolding $M_\infty$ is already contained in $M_{\mathrm{int}}$ as the principal block corresponding to lifted monomials that do not contain $t$. Therefore, $M_{\mathrm{int}}\succeq-\epsilon I$ automatically implies $M_\infty\succeq-\epsilon I$. This is different from the ETEP formulation in Section VII, where the interior and boundary tensor-eigenvalue problems are treated separately. After unfolding, we deal with an ordinary matrix eigenvalue problem. Compared with tensor eigenvalue computations, matrix eigenvalue algorithms are much more mature and numerically reliable, so the spectrum of $M_{\mathrm{int}}$ can be handled directly without separating the boundary problem.

For the low-order boundary, we mainly consider $j_0=2$, so that $A_{j_0}=A_2$ is a matrix and no unfolding relaxation is needed. This setting is also natural for controller design: even if the original system does not contain a linear term, a linear feedback term can be included in the controller to generate the quadratic term in $S(x)=-\dot V(x)$. We therefore impose
\begin{equation}
M_{\mathrm{int}}\succeq-\epsilon I,\qquad M_0=A_2\succeq0,\qquad \epsilon\geq0.
\label{eq:unfolding_psd_constraints}
\end{equation}
The first condition bounds the violation of the sufficient PSD unfolding condition by $\epsilon$, while the second condition treats the low-order quadratic boundary exactly.

If a linear control term is not available and the lowest nonzero even degree satisfies $j_0>2$, no additional near-PSD constraint on $M_0$ is required. Indeed, making the $S_{\mathrm{pd}}(x)$ in \eqref{eq:split_pd_psd} positive definite directly requires $\alpha A_{j_0}(\theta)x^{j_0}\geq \eta\|x\|_2^{j_0}$ with $\eta>0$, and therefore the lowest-order homogeneous part is already guaranteed to be positive definite.

Therefore, the proposed constraints should be interpreted as a tractable near-PSD relaxation rather than an exact characterization of the positivity of $S_{\mathrm{psd}}(x)$.

\subsection{ETEP-guided U-eigenvalue synthesis program}

Combining the PD anchor with the unfolding-based near-PSD relaxation, we obtain the following convex template:
\begin{equation}
\begin{aligned}
\max_{\theta,\eta,\varepsilon}\quad & \eta-\rho\varepsilon\\
\mathrm{s.t.}\quad
& \alpha A_{j_0}(\theta)x^{j_0}\ge \eta \|x\|_2^{j_0},\qquad \forall x\in \mathbb{R}^n,\\
& M_{\mathrm{int}}(\theta)\succeq -\varepsilon I,\\
& \varepsilon\ge 0,\\
& \theta\in \Theta,
\end{aligned}
\label{eq:ETEP_Ueig_program}
\end{equation}
where $\Theta$ denotes the admissible controller parameter set. The slack variable $\varepsilon\geq0$ is introduced to accommodate the conservatism of the unrestricted unfolding discussed above.  The parameter $\rho>0$ balances the desire for a large PD-anchor margin $\eta$ against a small near-PSD tolerance $\varepsilon$. Thus, the first constraint provides the strict low-order stabilizing margin, while the second searches for a higher-order part whose extended unfolding is as close as possible to the PSD cone.

When the design variable $\theta$ parameterizes directly the entries of the relevant coefficient tensors, or equivalently the entries of their unfolding matrices, $M_{\mathrm{int}}(\theta)$ depends affinely on $\theta$, since the unfolding operation is only a linear index rearrangement of tensor entries. For $j_0=2$, both the PD-anchor constraint and the near-PSD unfolding constraint are LMIs, and hence the entire program \eqref{eq:ETEP_Ueig_program} is a convex semidefinite program (SDP).

Compared with traditional sum-of-squares (SOS) synthesis, the proposed ETEP-guided U-eigenvalue route is particularly attractive when the problem already possesses an explicit tensor structure. Instead of introducing a large Gram matrix over a monomial basis, our method works directly with the coefficient tensors and their unfoldings. The low-order term provides an exact PD anchor, while $M_{\mathrm{int}}$ provides a computationally inexpensive spectral surrogate for the remaining higher-order part. 

From a computational viewpoint, this formulation is typically much lighter than SOS, whose cost is dominated by the size of the monomial basis and the associated Gram matrix, both of which grow rapidly with the dimension and polynomial degree. Hence, while SOS is more general and may provide a tighter positivity relaxation, the proposed method offers a scalable and transparent alternative for tensor-structured polynomial systems. In particular, it provides a cheap SDP-based controller synthesis procedure in which the strict low-order stability margin is retained exactly, while the higher-order polynomial terms are handled through a controlled near-PSD relaxation, providing a large domain of attraction.

\subsection{Guaranteed domain of attraction}

To show that the proposed near-PSD synthesis can indeed guarantee a sufficiently large domain of attraction, we derive below an explicit lower bound on its size. We focus on the practically relevant case $j_0=2$ and $k=4$. In this case,
$S_{\mathrm{pd}}(x)=\alpha A_2x^2$ and
$S_{\mathrm{psd}}(x)=(1-\alpha)A_2x^2+A_3x^3+A_4x^4$, where $\alpha\in(0,1)$. The homogenization of $S_{\mathrm{psd}}$ is
$S_{e,\mathrm{psd}}(x,t)=(1-\alpha)t^2A_2x^2+tA_3x^3+A_4x^4$.
Let $M_{\mathrm{int}}$ be its $(2,2)$-unfolding.

\begin{thm}[Guaranteed domain of attraction]
\label{thm:guaranteed_doa}
Suppose that the controller obtained from \eqref{eq:ETEP_Ueig_program} satisfies $\alpha A_2\succeq\eta I$ with $\eta>0$ and $M_{\mathrm{int}}\succeq-\varepsilon I$ with $\varepsilon\geq0$. Define $c_\alpha=(1+2\alpha)/(3\alpha)$. If $\varepsilon=0$, then the origin is globally asymptotically stable. If $\varepsilon>0$ and $c_\alpha\eta>2\varepsilon$, then
\begin{equation}
\mathcal B_R:=\{x\in\mathbb R^n:\|x\|_2<R\}\subseteq\mathcal D(0),
\qquad
R=\sqrt{\frac{c_\alpha\eta}{\varepsilon}-2},
\label{eq:guaranteed_doa_radius}
\end{equation}
where $\mathcal D(0)$ denotes the domain of attraction of the origin.
\end{thm}

\begin{proof}
Let $\bar z=(x,1)$ denote the homogenized variable evaluated at $t=1$, which recovers the original polynomial from its homogeneous extension, and define $v(\bar z)=\operatorname{vec}(\bar z^{\otimes2})$. Denote $y_0=v(e_t)$ and $w(x)=v(\bar z)-y_0$. Since $w(x)=v(\bar z)-y_0$, we have
$y_0^\top w(x)=y_0^\top v(\bar z)-y_0^\top y_0$. Moreover, $y_0=v(e_t)$ and $\bar z=(x,1)$ give $y_0^\top v(\bar z)=(e_t^\top\bar z)^2=1$ and $y_0^\top y_0=(e_t^\top e_t)^2=1$. Hence, $y_0^\top w(x)=0$. Moreover, $S_{\mathrm{psd}}(0)=0$, and therefore $y_0^\top M_{\mathrm{int}}y_0=0$. It follows that
$S_{\mathrm{psd}}(x)=2y_0^\top M_{\mathrm{int}}w(x)+w(x)^\top M_{\mathrm{int}}w(x)$.

From $M_{\mathrm{int}}\succeq-\varepsilon I$, we have
$w(x)^\top M_{\mathrm{int}}w(x)\geq-\varepsilon\|w(x)\|_2^2$. Moreover, $\|v(\bar z)\|_2^2=\|\bar z\|_2^4=(1+\|x\|_2^2)^2$. Since $y_0^\top w(x)=0$ and $\|y_0\|_2=1$, this gives
$\|w(x)\|_2^2=(1+\|x\|_2^2)^2-1=2\|x\|_2^2+\|x\|_2^4$.

It remains to evaluate the cross term $2y_0^\top M_{\mathrm{int}}w(x)$. Notice that the entries of the $t^2$ row associated with lifted coordinates of the form $x_it$ or $tx_i$ are zero, since $S_{e,\mathrm{psd}}$ contains no monomials of the form $t^3x_i$. Let $\tau$ denote the index associated with $t$. Under the supersymmetric homogenization of $(1-\alpha)t^2A_2x^2$, for each ordered pair $(i,j)$ the two $t$-indices can occupy $\binom{4}{2}=6$ positions, and hence
$(\mathcal T_{\mathrm{psd}})_{\tau\tau ij}
=\frac{1-\alpha}{6}(A_2)_{ij}.$
Consequently,
\[
2y_0^\top M_{\mathrm{int}}w(x)
=\frac{1-\alpha}{3}A_2x^2.
\]
The cubic and quartic terms do not contribute to this cross term since they contain only one and zero copies of $t$ respectively. Then,
$S_{\mathrm{psd}}(x)\geq\frac{1-\alpha}{3}A_2x^2-\varepsilon(2\|x\|_2^2+\|x\|_2^4)$.

Combining this inequality with $S_{\mathrm{pd}}(x)=\alpha A_2x^2$ gives
\begin{equation}
    S(x)\geq
\left(c_\alpha\eta-2\varepsilon-\varepsilon\|x\|_2^2\right)\|x\|_2^2,
\label{eq:doa_lower_bound}
\end{equation}
where we have used $\alpha A_2\succeq\eta I$, and hence $A_2x^2\geq(\eta/\alpha)\|x\|_2^2$.

If $\varepsilon=0$, \eqref{eq:doa_lower_bound} gives $S(x)\geq c_\alpha\eta\|x\|_2^2>0$ for every $x\neq0$. Since $S(x)=-\dot V(x)$ and $V(x)=\frac{1}{2}\|x\|_2^2$ is radially unbounded, the origin is globally asymptotically stable.

Now suppose $\varepsilon>0$. If $c_\alpha\eta>2\varepsilon$, the radius in \eqref{eq:guaranteed_doa_radius} is positive, and \eqref{eq:doa_lower_bound} implies $S(x)>0$ whenever $0<\|x\|_2<R$. To show that the whole open ball belongs to the domain of attraction, take any $x(0)$ satisfying $\|x(0)\|_2<R$ and choose $\bar R$ such that $\|x(0)\|_2<\bar R<R$. Then
$\delta:=c_\alpha\eta-2\varepsilon-\varepsilon\bar R^2>0$, and for every $\|x\|_2\leq\bar R$,
$S(x)\geq\delta\|x\|_2^2$. Hence $\dot V(x)\leq-2\delta V(x)$.

Since $V$ is decreasing, a trajectory starting inside $\mathcal B_{\bar R}$ cannot leave $\mathcal B_{\bar R}$. Therefore the above inequality holds for all $t\geq0$, and $V(x(t))\leq e^{-2\delta t}V(x(0))$. Thus $x(t)\to0$ as $t\to\infty$. Since this argument holds for every $x(0)\in\mathcal B_R$, we conclude that $\mathcal B_R\subseteq\mathcal D(0)$.
\end{proof}

Theorem \ref{thm:guaranteed_doa} gives a direct interpretation of the two quantities optimized in \eqref{eq:ETEP_Ueig_program}. In particular, the ratio $\eta/\varepsilon$ provides a direct measure of the trade-off between the stabilizing low-order component and the conservatism introduced by the unfolding relaxation.

We here consider $j_0=2$ and $k=4$, since this case is practically relevant and leads to a simple closed-form estimate. The same procedure extends to general even degree $k=2p$ by bounding the additional north-pole cross terms and deriving a scalar polynomial inequality in $r=\|x\|_2$. The guaranteed radius is then characterized by the largest positive root of the resulting polynomial inequality. For low-order cases this root may be available in closed form, whereas for sufficiently high polynomial degree no general radical expression exists, and the radius can instead be obtained from calculating the unique positive root numerically or from a conservative closed-form bound.

\begin{remark}
    Although the guaranteed domain-of-attraction bound in Theorem~\ref{thm:guaranteed_doa} depends directly on the ratio $\eta/\varepsilon$, we adopt the linear objective $\eta-\rho\varepsilon$. This choice preserves a standard convex SDP formulation, whereas directly optimizing the ratio would lead to a fractional, nonlinearly parameterized problem.
\end{remark}

\section{Illustrative Examples of the Tensor-eigenvalue Positivity Test}

This section presents 2 representative example to demonstrate the
practical use of the proposed tensor-eigenvalue framework for certifying
polynomial positivity.  The examples follows the  procedure:
\begin{enumerate}
\item compute all real solutions $((x,t),\lambda)$ of the extended
tensor-eigen system~\eqref{eq:EH-operator} (interior analysis);
\item evaluate the asymptotic boundary quantities
$\alpha_\infty$ and $\beta_0$ from~\eqref{eq:boundary-bounds};
\item determine the minimum
$\underline{\lambda}=\min\{\lambda_{\mathrm{int}},\alpha_\infty,\beta_0\}$
and conclude the definiteness of~$P$ according to
Theorem~\ref{thm:PD}.
\end{enumerate}

\subsection{Example 1: Nonhomogeneous quadratic–quartic system}

Consider
\[
P_{1}(x,y)=(x^2+y)^2=x^4+2x^2y+y^2,
\]
which contains a quartic, a cubic, and a quadratic term after expansion
in $x$ and $y$.
This example is nonhomogeneous and nonnegative for all $(x,y)\neq(0,0)$,
but standard H-eigenvalue tests fail because of the odd-order tensor.

\paragraph{Interior tensor-eigen computation.}
The homogeneous lifting is
\[
\widetilde P(x,t)
=x^4+2t x^2y+t^2y^2.
\]
Substituting into the extended tensor-eigen system~\eqref{eq:EH-operator}
gives
\[
\begin{cases}
4x^3+4t x y=\lambda x,\\[3pt]
2t x^2+2t^{\,2}y=\lambda y,\\[3pt]
2ty^2+2x^2 y=\lambda t,\\[3pt]
x^2+y^2+t^2=1,\ t>0.
\end{cases}
\]
As mentioned, it can be reformulated as a Z-eigenproblem.
Solving numerically via TenEig Matlab toolbox \cite{teneig} gives
$\lambda_{\mathrm{int}}=0$. \hma{We also calculate the minimum U-eigenvalue of the extended tensor and it is $-0.33333$. This shows that the U-eigenvalue is indeed a conservative bound for the Z-eigenvalue.}

\paragraph{Boundary evaluation.}
At $r\to\infty$, the dominant term is $x^4$, so
\[
\alpha_\infty=\inf_{\|u\|=1}u_1^4=0.
\]
At $r\to0^+$, the lowest even-order term is $y^2$, yielding
\[
\beta_0=\inf_{\|u\|=1}u_2^2=0.
\]

\paragraph{Positivity conclusion.}
Combining the results,
\[
\underline{\lambda}
=\min\{\lambda_{\mathrm{int}},\alpha_\infty,\beta_0\}
=\min\{0,0,0\}=0.
\]
Hence $P_{1}$ is \emph{positive semidefinite} with $\underline{\lambda}=0$.
The result verifies that the extended tensor-eigen framework successfully
captures positivity in nonhomogeneous coupled terms.

\subsection{Example 2: The Motzkin polynomial (non-SOS but PSD)}

The classical \emph{Motzkin polynomial} \cite{motzkin1967} is defined as
\[
M(x,y)=x^4y^2+x^2y^4+1-3x^2y^2.
\]
It is well known that $M(x,y)\ge0$ for all $(x,y)\in\mathbb{R}^2$,
yet it does not admit a sum-of-squares (SOS) representation.
This example demonstrates that the proposed
extended tensor–eigenvalue framework can nevertheless
certify its positivity.

\paragraph{Homogeneous lifting and extended eigen system.}
The highest degree of $M$ is $k=6$.
Introducing the homogenized form
\[
\widetilde M(x,y,t)=x^4y^2+x^2y^4+t^6-3x^2y^2t^2,
\]
the interior critical points satisfy the extended tensor–eigen system
\[
\begin{cases}
4x^3y^2+2xy^4-6xy^2t^2=\lambda x,\\[3pt]
2x^4y+4x^2y^3-6x^2yt^2=\lambda y,\\[3pt]
6t^5-6x^2y^2t=\lambda t,\\[3pt]
x^2+y^2+t^2=1,\quad t>0.
\end{cases}
\]
Solving this system numerically 
yields 
\[
\lambda_{\mathrm{int}} = 0.  
\]
\hma{We also calculate the minimum U-eigenvalue of the extended tensor and it is $-0.2$. This shows that the U-eigenvalue is indeed a conservative bound for the Z-eigenvalue.}

\paragraph{ Boundary behaviour.}
At $r\to\infty$, only the leading homogeneous terms dominate, giving
\[
\alpha_\infty
=\inf_{\|u\|=1}\left(u_1^4u_2^2+u_1^2u_2^4\right)=0.
\]
At $r\to0^+$, the constant term $1$ dominates, so
$\beta_0>0$.

\paragraph{Positivity conclusion.}
The combined spectral bound
\[
\underline{\lambda}
=\min\{\lambda_{\mathrm{int}},\alpha_\infty,\beta_0\}
=0
\]
yields $\underline{\lambda}\ge0$.
Hence, the Motzkin polynomial is \emph{positive semidefinite}
and correctly certified by our framework, which cannot be checked by normal SOS techniques.

\begin{remark}
Although $M(x,y)$ is not a sum of squares,
its positivity is captured analytically through
the extended tensor–eigen system \eqref{eq:EH-operator}.
This highlights the capability of the proposed spectral criterion
to certify the nonnegativity of
\emph{non-SOS polynomials}, going beyond
classical SOS-based semidefinite relaxations.
\end{remark}

\section{Illustrative example: controller synthesis via ETEP-guided U-eigenvalue relaxations and SOS}
\label{sec:example_control}

In this section, we illustrate the proposed controller design framework on a three-dimensional polynomial system, and compare the resulting ETEP-guided U-eigenvalue synthesis with a traditional sum-of-squares (SOS) based design. The comparison is carried out under the same open-loop system and the same polynomial controller parameterization.

\subsection{Open-loop system}

Consider the following open-loop polynomial system:
{\small\begin{equation}
\label{eq:example_open_loop}
\begin{aligned}
\dot x_1 &= 0.9x_1 + 0.7x_2 + 0.2x_3 + 0.6x_1x_2 - 0.25x_2x_3 + 0.18x_1^3,\\
\dot x_2 &= -0.3x_1 + 0.8x_2 + 0.5x_3 - 0.35x_1x_3 + 0.25x_1x_2 + 0.12x_2^3,\\
\dot x_3 &= 0.4x_1 - 0.2x_2 + 0.6x_3 + 0.3x_2x_3 - 0.2x_1x_2 + 0.10x_3^3.
\end{aligned}
\end{equation}}
The origin is not asymptotically stable for the open-loop system, and therefore feedback design is required.

\subsection{Polynomial controller parameterization}

We employ a cubic polynomial state-feedback controller
$u(x)=\begin{bmatrix}u_1(x)\\u_2(x)\\u_3(x)\end{bmatrix},$
where each component is parameterized by all monomials of degree up to three in $(x_1,x_2,x_3)$. Hence, $\Theta=[-20,20]^{57}$ to avoid high gain. Let $m(x)$ be a matrix stacking all the corresponding monomials so that
\begin{equation}
\label{eq:example_controller}
u_i(x)=\theta_i^\top m(x), \qquad i=1,2,3,
\end{equation}
where each $\theta_i\in\mathbb{R}^{19}$ is a vector of controller coefficients. Denoting the stacked controller parameter by
$\theta = \begin{bmatrix}\theta_1^\top & \theta_2^\top & \theta_3^\top\end{bmatrix}^\top \in \mathbb{R}^{57},$
the closed-loop system is
\begin{equation}
\label{eq:example_closed_loop}
\dot x = f(x)+u(x).
\end{equation}

As in Section~IX, we adopt the quadratic Lyapunov candidate
$V(x)=\frac{1}{2}x^\top x,$
and define
\begin{equation}
\label{eq:example_S}
S(x;\theta):=-\dot V(x)=-x^\top\bigl(f(x)+u(x)\bigr).
\end{equation}
The resulting polynomial is decomposed as
\[
S(x;\theta)=S_{\mathrm{pd}}(x;\theta)+S_{\mathrm{psd}}(x;\theta),
\]
where the quadratic part provides the low-order PD anchor, while the remaining higher-order part is treated through unfolding-based near-PSD relaxations.

\subsection{ETEP-guided U-eigenvalue synthesis}

For the ETEP-guided design, the quartic homogenized polynomial associated with $S_{\mathrm{psd}}$ is constructed, yielding the optimization program described in Section~IX. In particular, the design seeks to maximize the anchor margin while penalizing the near-PSD slack of the interior unfolding with $\rho=1, \alpha=0.7$:
$\max_{\theta,\eta,\varepsilon}\ \eta-\rho\varepsilon.$
To prevent the optimization from trading an excessively large violation of the unfolding-based PSD condition for a larger anchor margin, we additionally impose $0\leq\varepsilon\leq\bar\varepsilon$, where $\bar\varepsilon$ is a prescribed maximum admissible near-PSD tolerance. In this experiment, we set $\bar\varepsilon=0.01.$
In our implementation, the optimization was solved successfully. The obtained numerical results are
\begin{equation}
\label{eq:ueig_result}
\eta_{\mathrm{U\text{-}eig}} = 4.658019,\qquad
\varepsilon_{\mathrm{U\text{-}eig}} = 10^{-2},
\end{equation}
and the total optimization time was
\begin{equation}
\label{eq:ueig_time}
t_{\mathrm{U\text{-}eig}} = 0.063185\ \text{s}.
\end{equation}
These values indicate that the proposed relaxation is numerically tractable and yields a feasible  controller. Here, we use Matlab with sdpt3 solver. The whole procedure is summarized as the Algorithm \ref{alg:ueig_synthesis}.

\begin{algorithm}[t]
\caption{ETEP-guided U-eigenvalue controller synthesis}
\label{alg:ueig_synthesis}
\begin{algorithmic}[1]
\Require Polynomial vector field $f(x)$; controller basis $m(x)$; admissible set $\Theta$; splitting parameter $\alpha\in(0,1)$; penalty $\rho>0$.
\Ensure Controller coefficients $\theta^\star$.

\State Parameterize the controller as
$
u_i(x)=\theta_i^\top m(x),\qquad i=1,\ldots,n,
$
and stack all controller coefficients into $\theta$.

\State Form the closed-loop Lyapunov polynomial
$
S(x;\theta)
=
-x^\top\bigl(f(x)+u(x;\theta)\bigr).
$

\State Collect terms of equal degree and construct the corresponding supersymmetric coefficient tensors:
$
S(x;\theta)=\sum_{j=2}^{k} A_j(\theta)x^j .
$

\State Split
$
S_{\mathrm{pd}}(x;\theta)
=
\alpha A_2(\theta)x^2,
$
and
$
S_{\mathrm{psd}}(x;\theta)
=
(1-\alpha)A_2(\theta)x^2
+
\sum_{j=3}^{k}A_j(\theta)x^j .
$

\State Homogenize $S_{\mathrm{psd}}$ to degree $k$:
$
S_{e,\mathrm{psd}}(x,t;\theta)
=
(1-\alpha)t^{k-2}A_2(\theta)x^2
+
\sum_{j=3}^{k}
t^{k-j}A_j(\theta)x^j .
$

\State Construct the supersymmetric extended tensor
$T_{\mathrm{psd}}(\theta)$ satisfying
$
S_{e,\mathrm{psd}}(x,t;\theta)
=
T_{\mathrm{psd}}(\theta)z^k,
\qquad
z=(x,t).
$

\State Set $p=k/2$ and construct the $(p,p)$-unfolding
$
M_{\mathrm{int}}(\theta)
=
T_{\mathrm{psd}}^{(p)}(\theta).
$

\State Introduce the SDP variables $\theta$, $\eta$, and $\varepsilon$.

\State Solve
$
\left\{
\begin{aligned}
\max_{\theta,\eta,\varepsilon}\quad
& \eta-\rho\varepsilon\\
\mathrm{s.t.}\quad
& \alpha A_2(\theta)\succeq\eta I,\\
& M_{\mathrm{int}}(\theta)\succeq-\varepsilon I,\\
& \eta\ge0,\qquad 0\leq \varepsilon \leq \bar\varepsilon,\\
& \theta\in\Theta .
\end{aligned}
\right.
$

\State Recover the controller
$
u_i^\star(x)=(\theta_i^\star)^\top m(x),
\qquad i=1,\ldots,n.
$

\State \Return $\theta^\star$, $\eta^\star$, $\varepsilon^\star$, and $u^\star(x)$.
\end{algorithmic}
\end{algorithm}

\subsection{SOS-based controller synthesis}

For comparison, we also solve a standard SOS-based controller design problem under the same controller parameterization \eqref{eq:example_controller}. The idea is to enforce a polynomial Lyapunov decrease condition directly by requiring
\[
S(x;\theta)-\eta\|x\|_2^2
\]
to be a sum of squares. Since every SOS polynomial is globally nonnegative, this constraint provides a sufficient certificate for
\[
S(x;\theta)\ge \eta\|x\|_2^2>0,\qquad \forall x\neq 0,
\]
and hence guarantees asymptotic stability of the origin. In practice, this condition is implemented by introducing a Gram matrix representation of the polynomial and solving the resulting semidefinite program.

The SOS optimization was also solved successfully. The resulting margin is
\begin{equation}
\label{eq:sos_result}
\eta_{\mathrm{SOS}} = 19.10,
\end{equation}
and the total computational time is
\begin{equation}
\label{eq:sos_time}
t_{\mathrm{SOS}} = 0.750507\ \text{s}.
\end{equation}
Here, we use Matlab with sdpt3 solver. This standard SOS-based controller synthesis procedure used for comparison is breifly summarized in Algorithm~\ref{alg:sos_synthesis}.

\begin{algorithm}[t]
\caption{SOS-based controller synthesis}
\label{alg:sos_synthesis}
\begin{algorithmic}[1]
\Require $f(x)$, controller basis $m(x)$, admissible set $\Theta$.
\Ensure Controller coefficients $\theta^\star$.

\State Parameterize $u_i(x)=\theta_i^\top m(x)$ and form
$
S(x;\theta)=-x^\top\bigl(f(x)+u(x;\theta)\bigr).
$

\State Introduce $\eta\ge0$ and impose
$
S(x;\theta)-\eta\|x\|_2^2
=
q(x)^\top Qq(x),
\qquad Q\succeq0,
$
where $q(x)$ is a suitable monomial basis.

\State Match polynomial coefficients and solve
$
\max_{\theta,\eta,Q}\ \eta
\quad
\text{s.t.}\quad
Q\succeq0,\ \theta\in\Theta.
$

\State Recover $u^\star(x)$ from $\theta^\star$.

\State \Return $\theta^\star$, $\eta^\star$, and $u^\star(x)$.
\end{algorithmic}
\end{algorithm}

\subsection{Comparison of computational performance}

Both methods successfully produced feasible controller solutions. However, their computational costs differ significantly. The ETEP-guided U-eigenvalue synthesis required only
$0.063185\ \text{s},$
whereas the SOS-based design required
$0.750507\ \text{s}.$
Hence, in this example, the proposed ETEP-guided method is approximately
$\frac{0.750507}{0.063185}\approx 11.88$
times faster than the SOS-based approach.

On the other hand, a naive comparison gives
$\eta_{\mathrm{SOS}} = 19.10 > \eta_{\mathrm{U\text{-}eig}} = 4.658019.$
Although the two margins have different meanings, this comparison still provides a rough indication of the conservatism of the proposed relaxation.The present method sacrifices some tightness in exchange for substantially improved numerical efficiency. Moreover, the SOS design provides a strict global Lyapunov certificate, whereas the ETEP-guided design yields a positive PD-anchor margin with a small near-PSD slack and is further assessed through closed-loop simulations.

\subsection{Closed-loop trajectories under the ETEP-guided controller}

\begin{figure}[t]
    \centering
    \includegraphics[width=1\linewidth]{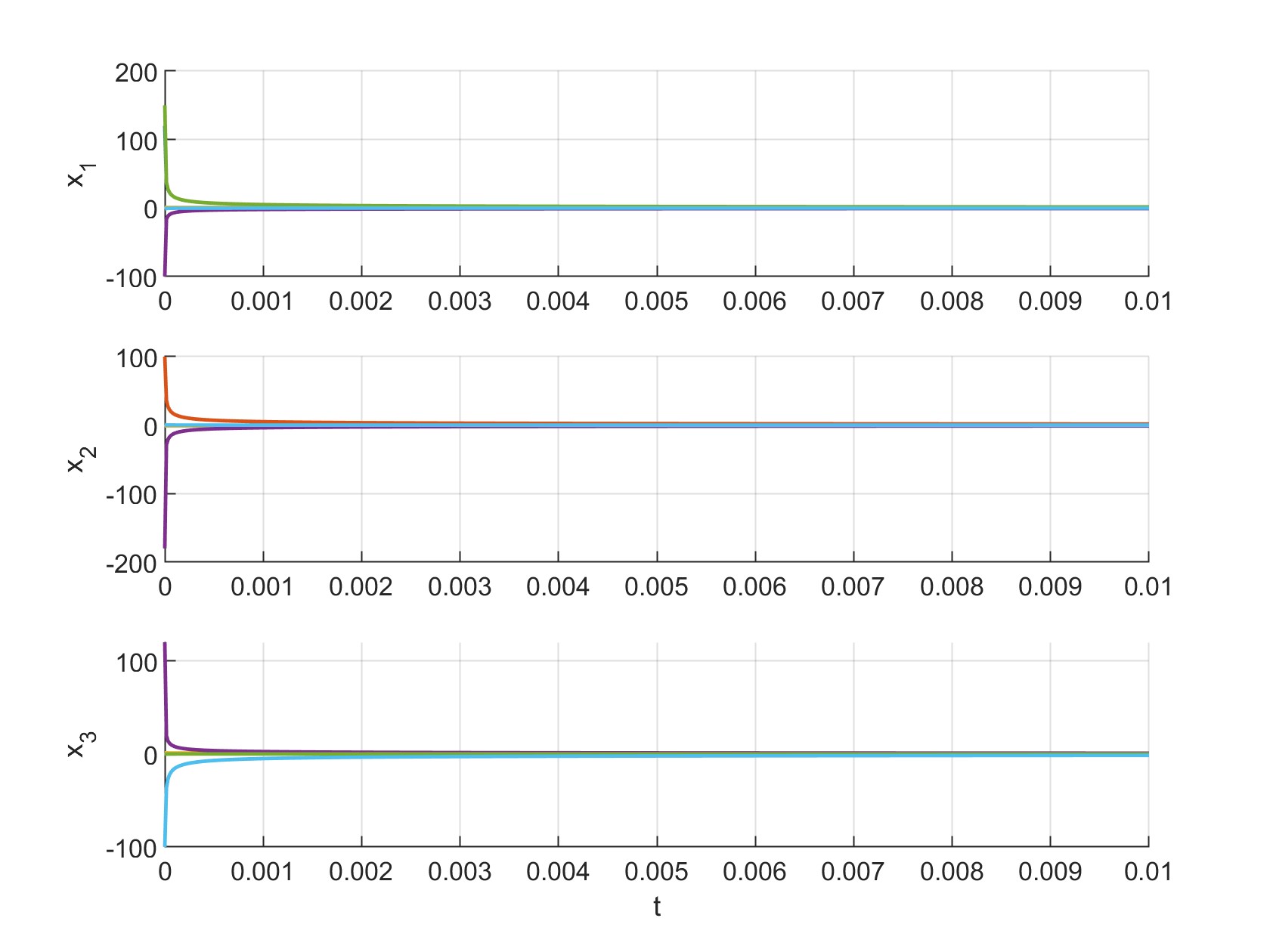}
    \caption{Closed-loop trajectories under the controller obtained by the ETEP-guided U-eigenvalue synthesis for multiple initial conditions. Different colors correspond to different initial conditions. Theorem~\ref{thm:guaranteed_doa} gives the guaranteed domain-of-attraction radius $R\approx23.03$. Some tested initial conditions have $|x_i(0)|>R$ and still converge to the origin, indicating that the theoretical radius is only a sufficient lower bound.}
    \label{fig:ueig_closedloop_traj}
\end{figure}

To further illustrate the effectiveness of the proposed method, Fig.~\ref{fig:ueig_closedloop_traj} shows the closed-loop trajectories generated by the controller obtained from the ETEP-guided U-eigenvalue synthesis, starting from multiple initial conditions. The trajectories converge toward the origin, confirming the stabilizing effect of the designed feedback law in simulation.

\section{Conclusion}
This paper developed a tensor-algebraic framework for the analysis and control of polynomial dynamical systems. By extending classical matrix-based ideas to higher-order tensor settings, we established systematic connections between tensor algebraic properties, Lyapunov stability, and spectral characterization for systems with multilinear and polynomial interactions.
We showed that positivity-related tensor structures, including positive definiteness, copositivity, and $P$-tensors, provide useful stability criteria for both homogeneous and nonhomogeneous polynomial systems. We further introduced an extended tensor--eigenvalue problem (ETEP) formulation, which gives a spectral--variational characterization of polynomial positivity and leads to a systematic stability test for polynomial vector fields. Building on this, we proposed an ETEP-guided U-eigenvalue synthesis framework based on tensor unfoldings, providing a structured and computationally tractable alternative to SOS-based controller design for tensor-structured polynomial systems.
Overall, the results extend the classical correspondence between algebraic structure and system dynamics from matrices to tensors, and provide a foundation for future studies on nonlinear systems with higher-order interactions.

\section*{Acknowledgement}
    Generative AI tools (ChatGPT) were used to assist with language and grammar editing and with programming support for the numerical implementation. All mathematical arguments, proofs, and algorithms were independently developed, checked, and verified by the authors. The authors take full responsibility for the content and correctness of the work.

\bibliographystyle{IEEEtran}
\bibliography{bib}

\begin{IEEEbiography}[{\includegraphics[width=1in,height=1.25in,clip,keepaspectratio]{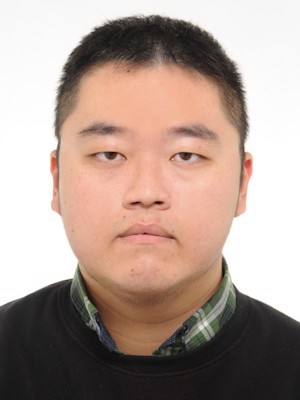}}]{Shaoxuan Cui}
received his B.Sc. degree in Electrical Engineering and Information Technology from the Technical University of Kaiserslautern, Germany, and Fuzhou University, China, in 2018, his M.Sc. degree in Electrical Engineering and Information Technology from the Technical University of Munich, Germany, in 2020, and his Ph.D. degree in Applied Mathematics from the University of Groningen, the Netherlands, in 2026.
He is currently a Postdoctoral Researcher with the Department of Applied Mathematics, The Hong Kong Polytechnic University, Hong Kong. His research interests include the modeling, analysis, and control of networked dynamical systems, with a particular focus on higher-order interactions.
\end{IEEEbiography}

\begin{IEEEbiography}
[{\includegraphics[width=1in,height=1.25in,clip,keepaspectratio]{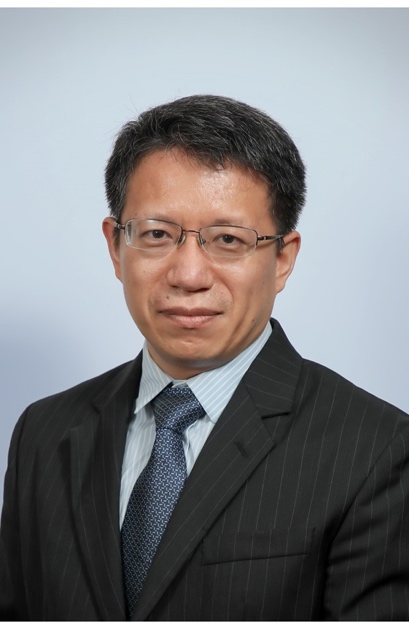}}]{Guofeng Zhang} received his B.Sc. degree and M.Sc. degree from Northeastern University, Shenyang, China, in 1998 and 2000 respectively. He received a Ph.D. degree in Applied Mathematics from the University of Alberta, Edmonton, Canada, in 2005. During 2005–2006, he was a Postdoc Fellow in the Department of Electrical and Computer Engineering at the University of Windsor, Windsor, Canada. He joined the School of Electronic Engineering of the University of Electronic Science and Technology of China, Chengdu, Sichuan, China, in 2007. From April 2010 to December 2011 he was a Research Fellow in the School of Engineering of The Australian National University. He joined the Hong Kong Polytechnic University in December 2011 and is currently a Professor in the Department of Applied Mathematics. His research interests include quantum information, quantum control, quantum algorithms, and tensor computation. He is an Associate Editor of IET Control Theory and Applications, Guest Associate Editor of International Journal of Bifurcation and Chaos, 2015 IEEE Multi-Conference on Systems and Control, Sydney, Australia, September 21-23, 2015, and the Manage Guest Editor for the Special Issue on Quantum Control and Quantum Machine Learning for Journal of The Franklin Institute. 
\end{IEEEbiography}

\begin{IEEEbiography}
[{\includegraphics[width=1in,height=1.25in,clip,keepaspectratio]{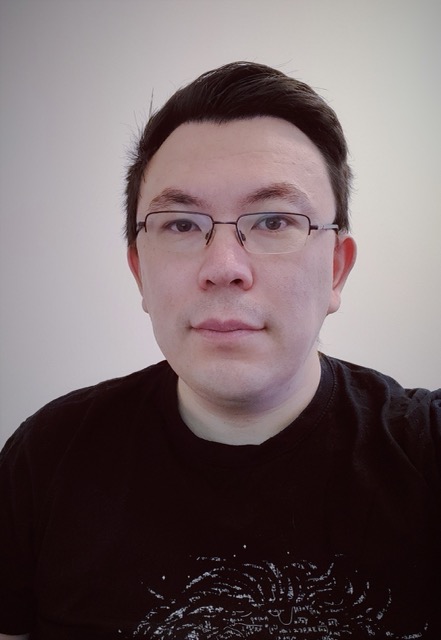}}]{Hildeberto Jardón-Kojakhmetov} is since 2020 an assistant professor with the group of Dynamical Systems, Geometry and Mathematical Physics at the University of Groningen, the Netherlands. He received his B.S. degree in Mechatronics Engineering from the National Polytechnic Institute (IPN), Mexico, in 2007 and the M.S. degree in Automatic Control from the Center for Research and Advanced Studies (CINVESTAV), Mexico, in 2010. In 2015 he received the Ph.D. degree in Mathematics from the University of Groningen, the Netherlands. From 2015 to 2018 he was a postdoctoral researcher at the University of Groningen, and from 2018 to 2020 a postdoctoral fellow at the Technical University of Munich. His research interests include slow-fast systems, singular perturbation problems, and network dynamics.
\end{IEEEbiography}

\begin{IEEEbiography}
[{\includegraphics[width=1in,height=1.25in,clip,keepaspectratio]{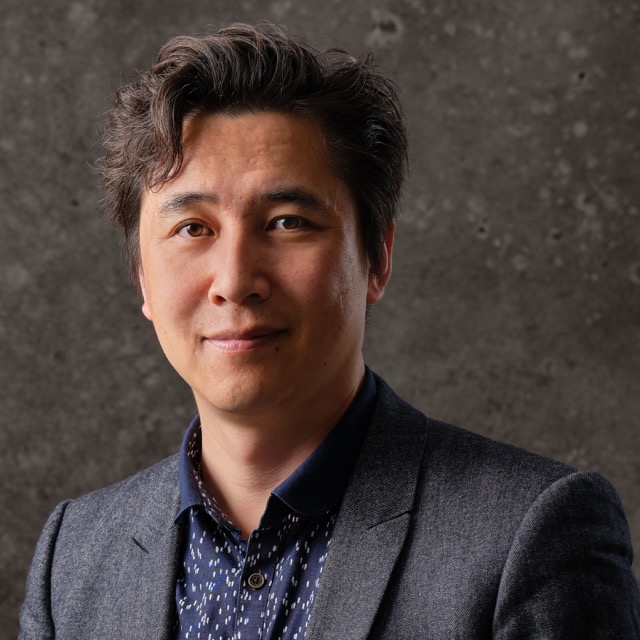}}]{Ming Cao} has since 2016 been a professor of networks and robotics with the Engineering and Technology Institute (ENTEG) at the University of Groningen, the Netherlands, where he started as an assistant professor in 2008. Since 2022 he is the director of the Jantina Tammes School of Digital Society, Technology and AI at the same university. He received the Bachelor degree in 1999 and the Master degree in 2002 from Tsinghua University, China, and the Ph.D. degree in 2007 from Yale University, USA. From 2007 to 2008, he was a Research Associate at Princeton University, USA. He worked as a research intern in 2006 at the IBM T. J. Watson Research Center, USA. He is the 2017 and inaugural recipient of the Manfred Thoma medal from the International Federation of Automatic Control (IFAC) and the 2016 recipient of the European Control Award sponsored by the European Control Association (EUCA). He is an IEEE fellow. He is a Senior Editor for Systems and Control Letters, an Associate Editor for IEEE Transactions on Automatic Control, IEEE Transaction of Control of Network Systems and IEEE Robotics and Automation Magazine, and was an associate editor for IEEE Transactions on Circuits and Systems and IEEE Circuits and Systems Magazine. He is a member of the IFAC Council and a vice chair of the IFAC Technical Committee on Large-Scale Complex Systems. His research interests include autonomous robots and multi-agent systems, complex networks and decision-making processes. 
\end{IEEEbiography}

\end{document}